\documentclass[aps,prx,preprint,superscriptaddress,longbibliography]{revtex4-2}

\usepackage{amsmath,amssymb}
\usepackage{graphicx}
\usepackage{bm}
\usepackage{hyperref}

\newcommand{\ve}[1]{\bm{#1}}

\begin{document}

\title{Observing the spatial and temporal evolution of exciton wave functions}

\author{Marcel Theilen}
\affiliation{Faculty of Physics and Materials Sciences Center, Philipps-Universit\"at Marburg, 35032 Marburg, Germany.}
\affiliation{Fachbereich Physik, University of Regensburg, 93053 Regensburg, Germany.}

\author{Siegfried Kaidisch}
\affiliation{Institute of Physics, University of Graz, 8010 Graz, Austria.}

\author{Monja Stettner}
\affiliation{Peter Gr\"unberg Institute (PGI-3), Forschungszentrum J\"ulich, 52425 J\"ulich, Germany.}
\affiliation{J\"ulich Aachen Research Alliance, Fundamentals of Future Information Technology, 52425 J\"ulich, Germany.}
\affiliation{Institut f\"ur Experimentalphysik IV A, RWTH Aachen University, 52074 Aachen, Germany.}

\author{Sarah Zajusch}
\affiliation{Faculty of Physics and Materials Sciences Center, Philipps-Universit\"at Marburg, 35032 Marburg, Germany.}

\author{Eric Fackelman}
\affiliation{Peter Gr\"unberg Institute (PGI-3), Forschungszentrum J\"ulich, 52425 J\"ulich, Germany.}
\affiliation{J\"ulich Aachen Research Alliance, Fundamentals of Future Information Technology, 52425 J\"ulich, Germany.}
\affiliation{Institut f\"ur Experimentalphysik IV A, RWTH Aachen University, 52074 Aachen, Germany.}

\author{Alexa Adamkiewicz}
\affiliation{Faculty of Physics and Materials Sciences Center, Philipps-Universit\"at Marburg, 35032 Marburg, Germany.}

\author{Robert Wallauer}
\affiliation{Faculty of Physics and Materials Sciences Center, Philipps-Universit\"at Marburg, 35032 Marburg, Germany.}

\author{Andreas Windischbacher}
\affiliation{Institute of Physics, University of Graz, 8010 Graz, Austria.}

\author{Christian S. Kern}
\affiliation{Institute of Physics, University of Graz, 8010 Graz, Austria.}

\author{Michael G. Ramsey}
\affiliation{Institute of Physics, University of Graz, 8010 Graz, Austria.}

\author{Fran\c{c}ois C. Bocquet}
\affiliation{Peter Gr\"unberg Institute (PGI-3), Forschungszentrum J\"ulich, 52425 J\"ulich, Germany.}
\affiliation{J\"ulich Aachen Research Alliance, Fundamentals of Future Information Technology, 52425 J\"ulich, Germany.}

\author{Serguei Soubatch}
\affiliation{Peter Gr\"unberg Institute (PGI-3), Forschungszentrum J\"ulich, 52425 J\"ulich, Germany.}
\affiliation{J\"ulich Aachen Research Alliance, Fundamentals of Future Information Technology, 52425 J\"ulich, Germany.}

\author{F. Stefan Tautz}
\affiliation{Peter Gr\"unberg Institute (PGI-3), Forschungszentrum J\"ulich, 52425 J\"ulich, Germany.}
\affiliation{J\"ulich Aachen Research Alliance, Fundamentals of Future Information Technology, 52425 J\"ulich, Germany.}
\affiliation{Institut f\"ur Experimentalphysik IV A, RWTH Aachen University, 52074 Aachen, Germany.}

\author{Ulrich H\"ofer}
\affiliation{Faculty of Physics and Materials Sciences Center, Philipps-Universit\"at Marburg, 35032 Marburg, Germany.}
\affiliation{Fachbereich Physik, University of Regensburg, 93053 Regensburg, Germany.}

\author{Peter Puschnig}
\affiliation{Institute of Physics, University of Graz, 8010 Graz, Austria.}

\date{\today}

\begin{abstract}
Excitons, the correlated electron-hole pairs governing optical and transport properties in organic semiconductors, have long resisted direct experimental access to their full quantum-mechanical wave functions. Here, we use femtosecond time-resolved photoemission orbital tomography (trPOT), combining high-harmonic probe pulses with time- and momentum-resolved photoelectron spectroscopy, to directly image the momentum-space distribution and ultrafast dynamics of excitons in $\alpha$-sexithiophene thin films. We introduce a quantitative model that enables reconstruction of the exciton wave function in real space, including both its spatial extent and its internal phase structure. The reconstructed wave function reveals coherent delocalization across approximately three molecular units and exhibits a characteristic phase modulation, consistent with ab initio calculations within the framework of many-body perturbation theory. Time-resolved measurements further show a $\sim 20$\% contraction of the exciton radius within 400 fs, providing direct evidence of self-trapping driven by exciton-phonon coupling. These results establish trPOT as a general and experimentally accessible approach for resolving exciton wave functions --- with spatial, phase, and temporal sensitivity --- in a broad class of molecular and low-dimensional materials.
\end{abstract}

\maketitle

\section{Introduction}

Excitons govern the optical response and energy transport in organic semiconductors \cite{Spano2006,Koehler2015}. While their energetic structure is well established, direct experimental access to the full quantum-mechanical exciton wave function, including its spatial envelope and internal phase, has remained out of reach. This gap is significant: the exciton wave function controls migration, dissociation, and coupling to the molecular lattice, and thus plays a central role in phenomena ranging from light harvesting \cite{Bronstein2011,Nelson2018,Sneyd2021} to quantum-coherent excitonics \cite{Scholes2023,Zhang2024,Yamauchi2024}.

Here, we introduce an approach that directly measures exciton wave functions with both spatial and phase resolution using femtosecond time-resolved photoemission orbital tomography (trPOT) \cite{Wallauer2020,Neef2023, Bennecke2024,Bennecke2025}. This technique combines pump-probe photoemission spectroscopy with momentum microscopy, yielding momentum-space distributions of the photoexcited state. We develop a quantitative and broadly applicable model that reconstructs the exciton wave function in real space from these momentum-space signatures. In contrast to prior studies in two-dimensional semiconductors \cite{Dong2021, Man2021, Schmitt2022, Karni2022, Bennecke2025,Mourzidis2025}, where only the width of the momentum distribution could be linked to the exciton radius, our method retrieves the internal structure of excitons in organic semiconductors, including phase modulations between molecular units.

The model reproduces key features obtained from ab initio many-body perturbation theory, by solving the qausi-particle equations within the GW approximation and the Bethe-Salpeter equation for correlated electron-hole pairs, enabling a direct comparison between experiment and theory. It also provides a framework for analyzing the temporal evolution of excitons. We observe a contraction of the exciton size by roughly 20\% within 400 fs, consistent with self-trapping driven by exciton-phonon coupling \cite{Roux2025}.

We use $\alpha$-sexithiophene (6T) thin films as a model system. 6T is a prototypical organic semiconductor, which is relevant for optoelectronic applications such as solar cells \cite{Dong2020}, and its photophysics has been extensively studied in the past, both, experimentally \cite{Frolov2000,Kouki2000,Varene2012,Bronsch2019} and theoretically \cite{Horst1999,Bussi2002,Leng2015}.
As substrate, we select oxygen-passivated Cu(110) which leads to the growth of uniaxially aligned molecules in a herringbone structure \cite{Sun2012}. This ensures intermolecular overlap of the molecules' $\pi$ electron systems, enabling exciton delocalization across multiple molecules, which is an essential condition to explore the crossover from localized Frenkel excitons to delocalized Frenkel-Wannier excitons \cite{Koehler2015,Chowdhury2025}. This contrasts with earlier trPOT work on weakly coupled molecules  \cite{Wallauer2020}, in which we have investigated flat-lying molecules with negligible intermolecular coupling, where excitations were essentially confined to single-molecule Frenkel excitons. In the present case, intermolecular overlap brings out the genuine excitonic structure, allowing us to access phase coherence and spatial delocalization across multiple molecular units.

\section{Results}

\subsection{Momentum mapping of the exciton}

\begin{figure} 
	\centering
	\includegraphics[width=12.1cm]{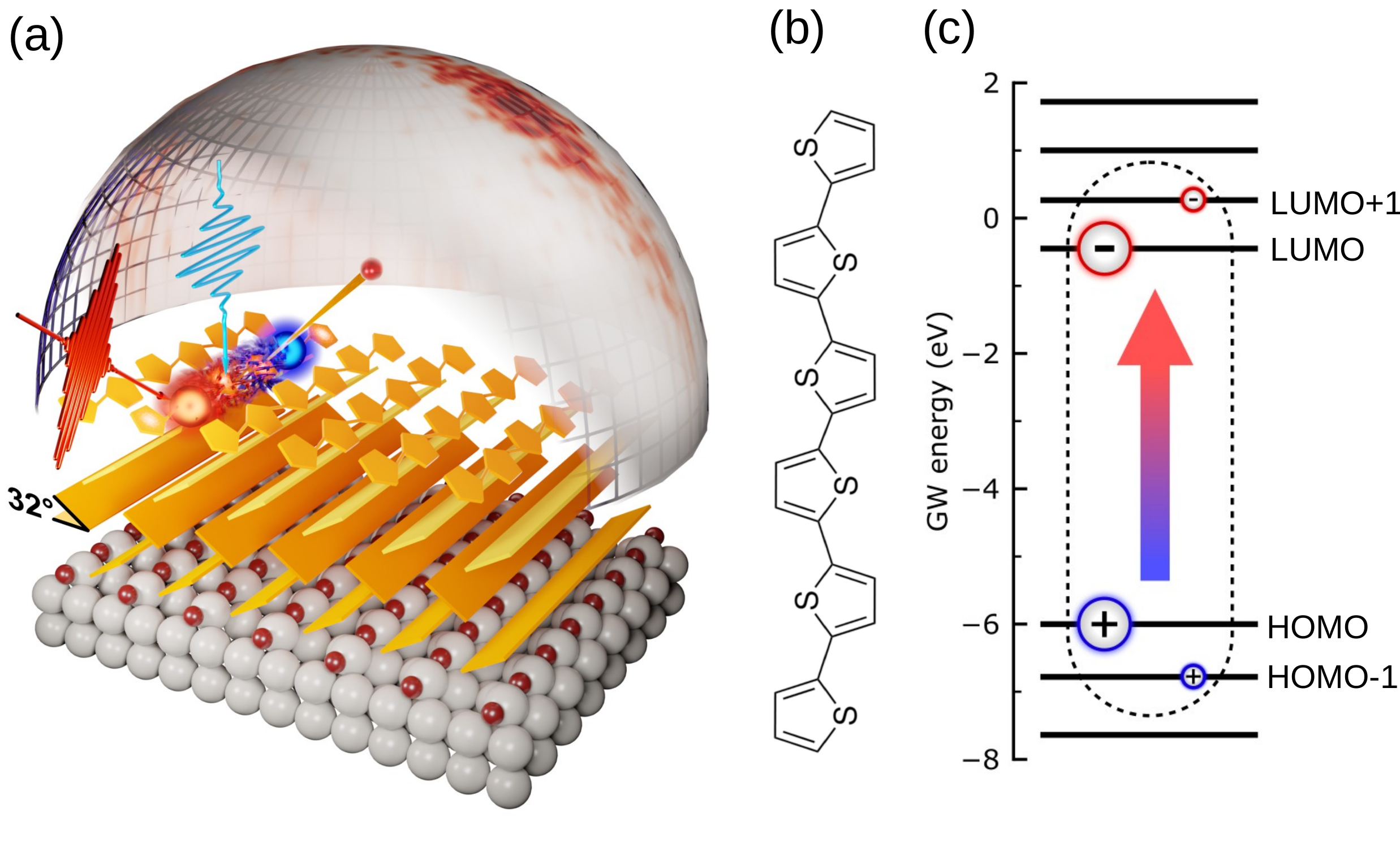} 
	\caption{(a) Structural model of four monolayers of 6T (yellow) on the Cu(110)-$(2 \times 1)$O substrate in which molecular planes are tilted by $\pm 32^\circ$ with respect to the sample surface (gray: copper atoms, red: oxygen atoms). The normal incidence pump (blue) and p-polarized probe at 55$^{\circ}$ incidence (red) pulses as well as the angular distribution of the emitted electrons are illustrated. 
		(b) Chemical structure of 6T. 
		(c) Illustration of the lowest optically allowed exciton state $S_1$ with the primary HOMO to LUMO and the secondary HOMO-1 to LUMO+1 contributions.}
	\label{fig1} 
\end{figure}

To directly observe the internal structure of an exciton, we performed ultrafast trPOT measurements on aligned $\alpha$-sexithiophene (6T) films, using femtosecond high-harmonic probe pulses, to record the momentum-space distribution and time evolution of the exciton.
As illustrated in Fig.~\ref{fig1}a, we have prepared a sample consisting of an oriented thin film of the rod-like organic semiconducting molecule 6T, whose chemical structure is depicted in Fig.~\ref{fig1}b, adsorbed on an oxygen-passivated Cu(110) surface. This substrate serves two important purposes. Firstly, it electronically decouples the molecules from the metallic substrate \cite{Wallauer2020}, thereby preventing a rapid quenching of the exciton. Secondly, the $(2 \times 1)$ missing-row O-reconstruction aligns the 6T molecules with their long axes along the $y=[001]$ direction of the Cu substrate since the molecules get trapped in the troughs between two adjacent Cu-O rows. Importantly, the molecules do not adsorb flat, but with their molecular planes tilted by $\sim 32^\circ$ about their long axis, as indicated in Fig.~\ref{fig1}a, thereby facilitating an effective overlap of the $\pi$ electron systems between neighboring molecules. For our experiments, we have chosen a four-monolayer film in which the molecules adopt a herringbone-like arrangement with alternating tilt angles, similarly to the stacking in the bulk crystal structure of 6T \cite{Horowitz1995}.

In order to select the photon energy and polarization direction of the pump laser pulse, we investigated the optical properties of 6T. In a first step, we performed calculations for an \emph{isolated} 6T molecule employing a many-body perturbation theory approach.
The resulting quasi-particle energies of the frontier molecular orbitals, obtained within the GW approximation, are depicted in Fig.~\ref{fig1}c and show a gap of 5.3~eV between the lowest unoccupied molecular orbital (LUMO) and the highest occupied molecular orbital (HOMO). The solution of the Bethe-Salpeter equation (BSE) then reveals the energetically lowest excited state $S_1$ to exhibit a transition dipole moment parallel to long axis of 6T at a transition energy of 2.8 eV. Thus, for an isolated 6T molecule a substantial exciton binding energy of 2.5 eV is to be expected. Note that the calculations also show that, in addition to the major HOMO to LUMO contribution, there is a secondary 8\%-share of a HOMO-1 to LUMO+1 transition in the  $S_1$ state. As shown in Refs.~\cite{Kern2023,Bennecke2025}, such an entangled excited state is predicted to exhibit a clear energy and momentum space signature. 

\begin{figure}[htb]
\centering
\includegraphics[width=\columnwidth]{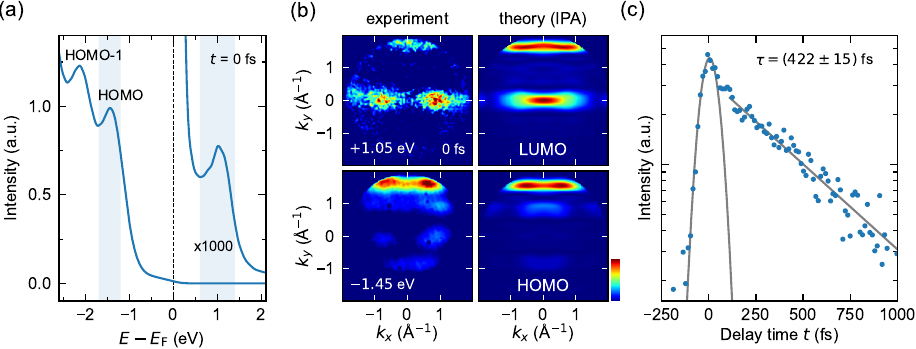} 
\caption{
 (a) Momentum-integrated photoemission spectrum with molecular emission from the HOMO and HOMO-1 below the Fermi energy $E_\mathrm{F}$ and the excitonic signature above $E_\mathrm{F}$ at $t=0$~fs. 
 (b) Momentum maps averaged over the blue-shaded energy regions indicated in panel (a) corresponding to the HOMO (lower row) and exciton (upper row). The experimental data (left column) are compared to simulations for the HOMO and LUMO, respectively, computed in the independent particle approximation (IPA) for a free-standing monolayer of 6T molecules. 
 (c) Momentum-integrated intensity, averaged over energies between 0.6 and 1.4~eV, as a function of delay time indicating an exponential decay of the exciton with a lifetime of  $\tau=(422 \pm 15)$~fs.}\label{fig2}
\end{figure}

Based on this analysis and on experimental optical reflection data \cite{Sun2012}, we used a normal-incidence pump pulse linearly polarized parallel to the long molecular axis for our trPOT experiments, in order to maximize the transition rate into the $S_1$ exciton. The pump energy was chosen to be 2.35 eV in line with earlier reflectance measurements on this 6T film \cite{Sun2012}, thus somewhat smaller than the computed transition energy for the isolated molecule.
The 21.7 eV probe pulse, created by high harmonics generation (HHG), was incident at an oblique angle with p-polarization, thereby optimizing the photoemission cross section for molecular emission. The photoelectrons were detected with a time-of-flight momentum microscope, allowing us to record a 4D data cube $I(E_\mathrm{kin}, k_x, k_y, t)$. Thus, the photoemission intensity $I$ was measured as a function of photoelectrons' kinetic energy $E_\mathrm{kin}$, their two parallel momentum components $k_x$ and $k_y$, and the delay time $t$ between the pump and the probe pulse, respectively. Note that in the actual analysis of the data, we reference energies to the Fermi energy $E_\mathrm{F}$, such that an energy value of zero corresponds to electrons emitted from the Fermi level, negative energies indicate occupied states while positive values signalize transiently excited states. Details of our experimental setup are described in Appendix~\ref{sec:trarpes}.

An overview of the photoemission data is given in Fig.~\ref{fig2}a, which shows momentum-integrated spectra at $t = 0$~fs both for occupied states below the Fermi energy $E_\mathrm{F}$ and for energies above, the latter enhanced by a factor of 1000 for the sake of clarity. Below $E_\mathrm{F}$, we find the HOMO and HOMO-1 emissions at about $-1.5$~eV and $-2.3$~eV, in agreement with a previous study \cite{Berkebile2009}. A momentum map taken around the HOMO peak is shown in the lower left corner of Fig.~\ref{fig2}b. The comparison with a simulation for a free-standing layer of 6T molecules (lower right corner) shows  excellent agreement and clearly supports this assignment. It is important to note that the simulation takes into account both the polarization of the probe pulse, which is incident from the negative $k_y$ direction and which gives rise to the ``up--down'' asymmetry in the maps, as well as the molecular tilt angle of $\pm 32^\circ$ in the molecular film which has been determined by static POT \cite{Puschnig2009a}.

Above $E_\mathrm{F}$, the pump pulse creates an excitation whose photoemission signature peaks at $1.05$~eV. As shown by the delay scan in Fig.~\ref{fig2}c, we observe a single-exponential decay with a lifetime of $(422 \pm 15)$~fs, which is somewhat larger than the 250~fs observed for a single layer of PTCDA molecules on the oxygen-passivated Cu(110) surface \cite{Wallauer2020}, presumably owing to the thicker organic film in the present case.
The momentum distribution at $t=0$~fs is shown in the top left corner of Fig.~\ref{fig2}b. 
It is characterized by two emission features (major lobes) along the $k_y=0$ line symmetrically grouped around the $\bar \Gamma$ point, and another weaker, minor lobe at $k_y\approx 1.7$~{\AA}$^{-1}$.
In a first attempt, we associate this observed momentum pattern with the LUMO of 6T,
which is depicted in the top right corner of Fig.~\ref{fig2}b. 
For simulating this map, we assumed that an electron populates the LUMO band of the free-standing 6T layer, but no interactions whatsoever with the remaining hole are considered. Such an approach has proven successful for PTCDA on oxygen-passivated Cu \cite{Wallauer2020}.
In contrast, the experimental map for 6T shows marked differences to the so-obtained simulated map. While the map employing the independent particle approximation (IPA) is characterized by two stripe-like features along $k_x$ in the $k$ regions of the major and minor lobes, respectively, the simulation fails to reproduce the observed intensity modulation along the $k_x$ direction in the major lobe region, resulting in an intensity maximum at normal emission, rather than two separated peaks  with a minimum at normal emission in the measurement. Moreover, also the minor lobe appears more extended in the simulation, while it is concentrated around $k_x=0$ in the experimental map. 
A possible origin for this discrepancy could be that electron-hole correlations cannot be ignored for 6T. 
Furthermore, it needs to be clarified whether an additional LUMO+1 contribution, as predicted for an isolated 6T molecule, could contribute to the deviation between  experiment and theory regarding the momentum distribution of the excited state.

\subsection{Exciton model}

\begin{figure}[htb]
\centering
\includegraphics[width=0.68\columnwidth]{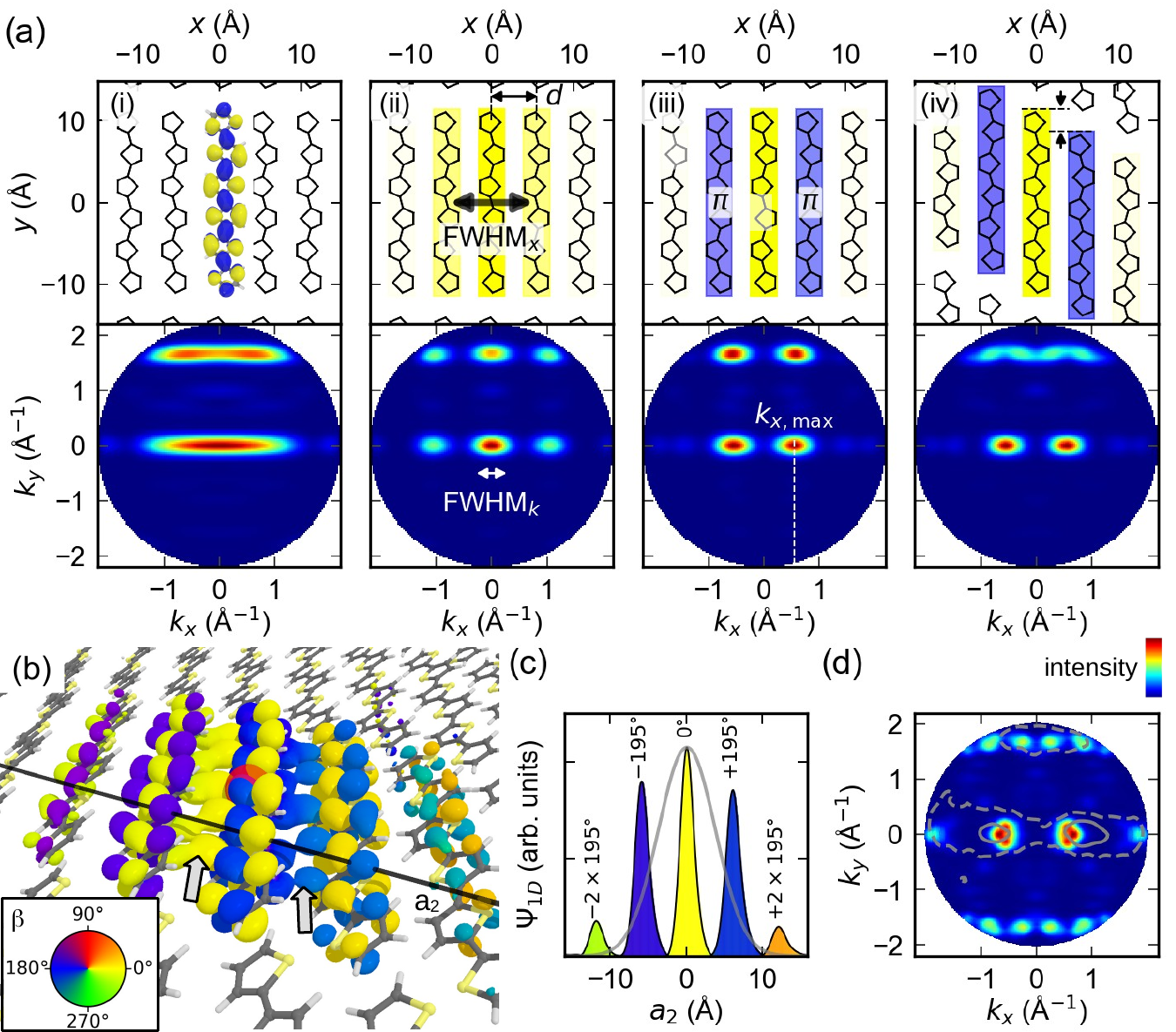}
\caption{ 
(a) Illustration of the exciton model in real space (upper row) and momentum space (lower row) according to Eqs.~\ref{eq:excitonmodel1} and \ref{eq:excitonmodel2}, respectively. The four columns (i--iv) correspond to different choices for the envelope function $\alpha(\ve{R})$, indicated by the black arrow, the phase modulation $e^{\mathrm{i} \beta(\ve{R})}$, and the unit cell shape as detailed in the text. 
(b) Electron distribution $\Psi(\ve{r}_e, \ve{r}_h)$ of the \emph{ab initio} exciton $S_1$ with respect to a fixed hole position $\ve{r}_h$ indicated by the red sphere located at the central molecule. For the visualization of the isosurface, we have chosen a value  of  10\% (3.5\%) of the maximum value of the absolute value $|\Psi|$ for the central three (outer two) molecules, while the phase $\arg (\Psi)$ is color-coded (see color-wheel in the inset). 
(c) One-dimensional line cut of the complex wave function $\Psi(\ve{r}_e, \ve{r}_h=\mathrm{fixed})$ along crystalline $\ve{a}_2$ direction (black line in panel (b)).  The solid line represents the absolute value and the filled colors the phase $\beta$, respectively. The gray line shows a Gaussian envelope extracted from the experimental map. 
(d) Momentum map corresponding to exciton $S_1$ according to Eq.~\ref{eq:exPOTperiodicsimple}. The solid (dashed) gray lines are isolines of the experimental exciton map (see Fig.~\ref{fig2}b) at 50\% (5\%) of its intensity maximum.}\label{fig3}
\end{figure}

To analyze the experimental data in more detail, we employ a model exciton wave function that captures all essential physics and which allows us to extract key properties of the exciton from experimental momentum maps. To reveal the spatial extent of the exciton beyond a single molecule, we employ a linear combination of LUMOs $\xi_L$ (of an isolated, single 6T molecule) at molecular sites $\ve{R}$ modulated by a real-valued envelope function $\alpha(\ve{R})$ and a complex, site-dependent phase factor $e^{\mathrm{i} \beta(\ve{R})}$. This leads to the following expression which describes the probability amplitude $\psi(\ve{r}_e)$ of finding an electron with respect to a fixed hole state (see Appendix~\ref{subsec:wannierDyson})
\begin{equation}
\label{eq:excitonmodel1}
\psi(\ve{r}_e) = \sum_{\ve{R}} \alpha(\ve{R}) e^{\mathrm{i} \beta(\ve{R})} \xi_L(\ve{r}_e - \ve{R}).
\end{equation}
Following the approach of photoemission orbital tomography, and as detailed in Appendix~\ref{subsec:derivemodel}, we compute photoemission momentum intensity maps, $I_\mathrm{model}(\ve{k})$, from the absolute value squared of the Fourier transform of $\psi(\ve{r}_e)$ defined in Eq.~\ref{eq:excitonmodel1}. It is straightforward to see that $I_\mathrm{model}(\ve{k})$ can then be expressed as the product of the Fourier transform $\mathcal{F} \left[ \xi_L \right] (\ve{k})$ of the LUMO with the Fourier series of the complex envelope function in the following way
\begin{equation}
\label{eq:excitonmodel2}
I_\mathrm{model}(\ve{k}) \propto \left| \sum_{\ve{R}} \alpha(\ve{R}) e^{\mathrm{i} \beta(\ve{R})} e^{-\mathrm{i} \ve{k} \cdot\ve{R}} \right|^2 \cdot \left| \mathcal{F} \left[ \xi_L \right] (\ve{k}) \right|^2.
\end{equation} 
The outcome of Eq.~\ref{eq:excitonmodel2} is illustrated in Fig.~\ref{fig3}a for various choices of the envelope function $\alpha(\ve{R})$, the phase modulation $e^{\mathrm{i} \beta(\ve{R})}$, and the unit cell shape. We begin with the limiting case in which the electron is confined to a single molecule, thus $\alpha(\ve{R})$ is sharply peaked at the central molecule, $\ve{R}=0$, and does not extend to adjacent molecules. In this Frenkel-exciton limit shown in column I, the sum over $\ve{R}$ collapses to a single contribution and the resulting momentum map reduces to the one of an isolated 6T molecular LUMO, $\left| \mathcal{F} \left[ \xi_L \right] (\ve{k}) \right|^2$, shown in the top left panel of Fig.~\ref{fig3}a. It results in the two characteristic stripe-like emission patterns discussed above. However, when allowing the electron to extend to neighboring molecules, for illustration we set $\alpha(\ve{R})$ to a Gaussian function with a full width at half maximum ($\mathrm{FWHM}_x$) indicated by the black arrow in (ii), the stripes are modulated in a characteristic fashion. The width of the intensity maximum ($\mathrm{FWHM}_k$) at the $\bar{\Gamma}$ point, shown by the white arrow, turns out to be inversely proportional to the real-space extension of the exciton. 

Note that in (ii), a possible phase modulation has been ignored by setting $\beta(\ve{R})=0$, resulting in an intensity maximum at $\bar{\Gamma}$ in disagreement with the experimental observations. Therefore in (iii), we choose $\beta(\ve{R})$ such that the wave function acquires a phase flip of $\pi$ between the LUMOs of adjacent molecules along the $x$ direction. Thereby, the intensity maxima are shifted in momentum space. Along the $k_y=0$ line, now two intensity maxima at $k_{x,\max} = \pm \frac{\pi}{d}\approx \pm 0.6$~{\AA}$^{-1}$ appear, where {$d \approx 5.5$~{\AA}, \cite{Horowitz1995}} denotes the lateral spacing of two adjacent molecules in $x$ direction. Thus, our model already accounts for the measured intensity modulations in the major lobe region. From the experimental map (Fig.~\ref{fig2}b), we extract peak maxima at $k_{x,\max} \approx \pm 0.85$~{\AA}$^{-1}$, in reasonable agreement with the model, and a $\mathrm{FWHM}_k$ of 0.63~{\AA}$^{-1}$. Using the relation $\mathrm{FWHM}_k \times \mathrm{FWHM}_x = 8 \ln 2$ for a Gaussian, the latter translates into a real-space $\mathrm{FWHM}_x$ of approximately $9$~{\AA}, thus suggesting an exciton which clearly extends over about three molecules. 

Upon closer inspection of the minor lobe region for the model result (iii), however, a minimum at $k_x=0$ is observed, while the experimental map shows a maximum. Owing to the lattice sum in Eq.~(\ref{eq:excitonmodel2}), the exact position of intensity extrema in the minor lobe region is sensitive to geometric details of the unit cell. In fact, choosing a vertical ``slip'' of 2.8~{\AA} between neighboring molecules (indicated by the black arrows in panel (iv)), which is consistent with the bulk 6T crystal structure \cite{Horowitz1995}, also improves the experimental agreement of the model result (iv) in this minor lobe region with essentially no changes in the major lobe region.

\subsection{\emph{Ab initio} calculations}

Given the success of our intuitive exciton model, we wanted to put it on firmer grounds and therefore performed \emph{ab initio} GW/BSE calculations beyond the isolated molecule limit. In particular, we treated a cluster of four 6T molecules and a periodic, free-standing monolayer of 6T molecules. The computational details of these two approaches are described in Appendices~\ref{sec:cluster} and \ref{sec:periodic}, respectively. 
Our first observation from these \emph{ab initio} computations regards the weaker LUMO+1 share to the first exciton. While for an isolated 6T molecule this contribution is $8 \%$, it reduces to $6 \%$ in the tetramer and amounts to only $3\%$ for the extended 6T layer. \emph{A posteriori}, this justifies the simplified model ansatz used in Eq.~\ref{eq:excitonmodel1}, in which only the LUMO has been considered. It also explains why experimentally no signatures of the LUMO+1 could be detected in the pump-probe photoemission data. 
(Note that given the fact that electronic LUMO+1 contributions to the exciton wave function are connected to HOMO-1 contributions of the hole, energy conservation demands \cite{Kern2023} the LUMO+1 emission features to be expected 0.8 eV below the peak at $E-E_\mathrm{F}\approx 1.05$~eV (compare Fig.~\ref{fig2}a). Taking into account the rapidly rising photoemission background when approaching the Fermi level and the predicted approximately 20--30 times lower intensity than the main peak, a possible LUMO+1 contribution is therefore extremely hard to detect experimentally.)

The second finding from our \emph{ab initio} GW/BSE calculations strengthens our model ansatz by connecting the spatial structure of the exciton wave function with certain characteristics of the  momentum map. For that purpose, we focus on our results for the free-standing 6T layer and the state $S_1$, the major contribution of the primary absorption peak (compare Supplemental Fig.~\ref{sifig3}). 
A common way to picture an electron-hole wave function $\Psi(\ve{r}_e, \ve{r}_h)$ is to plot the probability amplitude of finding an electron with respect to a hole fixed in space. The \emph{ab initio} exciton wave function for $S_1$ is shown in Fig.~\ref{fig3}b, where the hole, visualized by the red sphere, has been positioned above a carbon atom connecting the central thiophene rings. 
Rather than showing just the absolute value $\left| \Psi(\ve{r}_e, \ve{r}_h=\mathrm{const}) \right|$, which we represent by an isosurface at 10\% of the maximum value, we also visualize the color-coded phase of $\Psi$, an essential factor following our discussion of the model above. The wave function depicted in Fig.~\ref{fig3}b allows us to make three observations. 
Firstly, the exciton is indeed not constrained to a single molecule, but extends over several molecules. From the line cut along the $a_2$ stacking direction (compare unit cell in Supplemental Fig.~\ref{sifig1}), we extract an envelope function resulting in a $\mathrm{FWHM}_x$ of the probability amplitude of $\sim 16$~{\AA} somewhat larger than the experimental value shown as a gray line in Fig.~\ref{fig3}c. 
Secondly, we can clearly recognize the nodal patterns of the LUMO (compare Fig.~\ref{fig3}(a,i)). 
Thirdly, we clearly notice a characteristic phase variation across adjacent molecules that can best be seen by inspecting the line cut through the upper half of the molecular $\pi$ orbitals shown in Fig.~\ref{fig3}c. Specifically, the phase changes by $\Delta \beta \approx 195^\circ$ when moving from the central molecule to the right or left, respectively, thus by a value close to the phase flip of $\pi$ assumed in panels (iii) and (iv) of Fig.~\ref{fig3}a in our model exciton wave function.
Note that this phase relationship between adjacent molecules' LUMO orbitals results in a bonding intermolecular combination at their region of overlap indicated by the arrows in Fig.~\ref{fig3}b. 
It is worth mentioning that our GW/BSE calculations also predict an energetically higher lying, also optically active, $S_3$ state with a different phase relation that leads to an antibonding  intermolecular overlap between the three central molecules (compare Supplemental Fig.~\ref{sifig2}).

To obtain a momentum map of the $S_1$ exciton calculated from the \emph{ab initio} exciton wave function, we have extended the exPOT approach \cite{Kern2023} to periodic systems, allowing us to address excitons in the extended 6T layer studied here (see Appendix~\ref{sec:periodic}). Owing to the negligible contributions of unoccupied states above the LUMO to $S_1$, the general expression for photoemission intensity $I(\ve{k})$, see Eq.~\ref{eq:exPOTperiodic}, reduces to a simple product between the eigenvector of the effective electron-hole BSE Hamiltonian, $X_{v_1c_1\ve{k}}^{S_1}$, and the Fourier transform of the lowest conduction band orbital $\mathcal{F}  \left[ \xi_{c_1\ve{k}} \right]$, see Eq.~\ref{eq:expotperiodic_6T2},
\begin{equation}
\label{eq:exPOTperiodicsimple}
I_{S_1}(\ve{k}) \propto  \left| X_{v_1c_1\ve{k}}^{S_1} \right|^2 \cdot \left| \mathcal{F}  \left[ \xi_{c_1\ve{k}} \right] \right|^2.
\end{equation} 
Thus, the overall structure of this \emph{ab initio} expression for the photoemission intensity closely resembles the one of the model exciton wave function, Eq.~\ref{eq:excitonmodel2}, further justifying the model. The Fourier transform of the LUMO $\mathcal{F} \left[ \xi_L \right]$ should be identified with the Fourier transform of the lowest conduction state $c_1$, $\mathcal{F}  \left[ \xi_{c_1\ve{k}} \right]$, and the lattice sum appearing in Eq.~\ref{eq:excitonmodel2} corresponds to the momentum space representation of the BSE eigenvector $X_{v_1c_1\ve{k}}^{S_1}$ for the respective exciton $S_1$. 
Details on the relation between the exciton model and the first-principles description can be found in Appendix~\ref{sec:modelvsabinitio}, where direct comparisons of the model predictions with the \emph{ab initio} results are shown in Figs.~\ref{sifig5} and \ref{sifig6} for excitons $S_1$ and $S_3$, respectively.

The map resulting from the \emph{ab initio} approach, Eq.~\ref{eq:exPOTperiodicsimple}, is depicted in Fig.~\ref{fig3}d and agrees well with both the model result (iv) and the experimental momentum map whose main intensity features are included as gray isolines in Fig.~\ref{fig3}d. 
In particular, also the \emph{ab initio} map exhibits pronounced intensity maxima along the major lobe region, whose widths $\mathrm{FWHM}_k$ are consistent with the $\mathrm{FWHM}_x$ of the envelope of the exciton and whose $k_x$-positions are related to the above-mentioned phase modulation with $\Delta \beta \approx 195^\circ$. Using $\lambda = \frac{360^\circ}{\Delta \beta} d$, with $d$ denoting the molecular spacing (see Supplemental Fig.~\ref{sifig1}), explains the appearance of the maxima at $k_{x,\max} = \pm \frac{2 \pi}{\lambda} \approx \pm 0.64$~{\AA}$^{-1}$.

\subsection{Time evolution of exciton}

\begin{figure}[h]
\centering
\includegraphics[width=0.9\columnwidth]{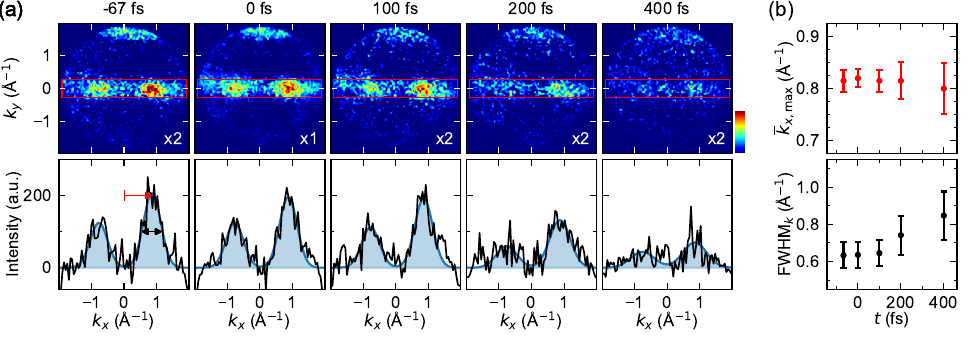}
\caption{ 
(a) Experimental momentum maps at five different pump/probe delay times $t$ (upper row) together with intensity linescans (lower row) along $k_x$ in the red-boxed regions. The blue-shaded areas correspond to the fits to the data (black line) using two independent Gaussians with a common $\mathrm{FWHM}_k$.  
(b) Position (red symbols) and full width at half maximum (black symbols) of the intensity peaks as a function of delay time resulting from the fits shown in panel (a).}\label{fig4}
\end{figure}

Having a validated model at hand, we are in a position to reliably extract the temporal evolution of the exciton wave function from the experimentally observed delay-time dependence of the momentum signatures depicted in Fig.~\ref{fig4}. Panel (a) shows experimental momentum maps of the $S_1$ exciton at $+1.05$~eV above $E_\mathrm{F}$ for five different delay times. We now analyze the major lobe region (red boxes) in more detail by plotting intensity linescans along $k_x$ shown as black lines in the lower row of Fig.~\ref{fig4}a. 
According to our exciton model, we fit the linescans by two Gaussians with in total five adjustable parameters, namely one common $\mathrm{FWHM}_k$ and two independent peak positions ($k_{x, \text{max}}^{-}$ and $k_{x, \text{max}}^{+}$, with Fig.~\ref{fig4}b showing $\bar{k}_{x, \text{max}} = \frac{1}{2} (|k_{x, \text{max}}^{-}| + k_{x, \text{max}}^{+})$) and peak heights to account for the asymmetry in the experimental data which arises from a slight misalignment of the sample.
The resulting best fits and the statistical uncertainties of the fit parameters are depicted in Figs.~\ref{fig4}A and B, respectively. 

Firstly, we observe a significant increase of the $\mathrm{FWHM}_k$ from $ 0.63$~{\AA}$^{-1}$ at $t = 0$~fs to about $0.8$~{\AA}$^{-1}$ at 400~fs, which translates into a localization of the real space exciton envelope from initial 8.8~{\AA} to 6.9~{\AA}. Such a localization, which is expected to be accompanied by a slight increase in the exciton binding energy, in consistency with our experimental data (see Appendix~\ref{sec:energyspectrum}), may arise from polaronic effects due to exciton-phonon coupling. 
A comprehensive microscopic theory of exciton-phonon coupling that could account for our experimental findings lies beyond the scope of this study. Nevertheless, the combination of time-resolved momentum imaging and our model introduces a new set of measurable observables that offer a solid benchmark for future theoretical work.
Secondly, within  statistical uncertainties, the position of the intensity maxima $k_{x,\max}$ remains nearly constant over time, reflecting the fact that the phase modulation retains the bonding character during the entire lifetime of the exciton.
Together, these findings establish a direct experimental view of both the spatial extent and internal phase coherence of excitons, and how these evolve under the influence of exciton-phonon coupling.

\section{Conclusion}

We have introduced a general method for reconstructing exciton wave functions from time-resolved momentum-resolved photoemission measurements. Applied to $\alpha$-sexithiophene, this approach reveals both the spatial delocalization and internal phase structure of the exciton, in quantitative agreement with GW/BSE calculations. The observed coherent extension across three molecular units and the characteristic phase modulation demonstrate the mixed Frenkel-Wannier nature of the exciton in this system.
Time-resolved measurements further show a $\sim 20$\% reduction in exciton size within 400 fs, providing direct evidence of self-trapping facilitated by exciton-phonon coupling. This capability to track both the spatial and temporal evolution of the exciton wave function represents a substantial advance beyond envelope-function imaging in two-dimensional materials.
Because the method relies on general features of the photoemission process and does not assume system-specific simplifications, it provides a widely applicable framework for visualizing excitonic correlations in molecular, hybrid, and low-dimensional materials. It thus opens new opportunities for probing many-body quasiparticle dynamics and for benchmarking theoretical approaches to correlated electron-hole pairs.

\begin{acknowledgments}
The computational results have been achieved using the Austrian Scientific Computing (ASC) infrastructure.
\paragraph*{Funding:}
We acknowledge support from the European Research Council (ERC), Synergy Grant ``Orbital Cinema'', Project ID 101071259.
\paragraph*{Author contributions:}
F.S.T., U.H, and P.P. conceived and designed the research. M.S. and E.F. prepared the $\alpha$-sexithiophene samples with support from F.C.B. and S.S.. 
M.T. and M.S. performed the time-resolved photoemission (trARPES) experiments; E.F., F.C.B., and A.A. assisted in carrying them out.
S.Z., R.W. and U.H. designed and constructed the experimental setup.
S.K., with support from C.K. and A.W., performed the \emph{ab initio} GW/BSE calculations. S.K. analyzed all theoretical data and prepared the corresponding figures.
M.T. analyzed all experimental data and prepared the time-resolved POT figures.
A.W. and M.G.R. proposed the studied system.
F.S.T., U.H., and P.P. supervised the Ph.D. work of M.S., M.T., and S.K., respectively.
P.P. wrote the manuscript with significant input from M.T., S.K., M.G.R., F.S.T., and U.H. 
All authors discussed the results, contributed to the interpretation, and reviewed and improved the final manuscript and figures.
\paragraph*{Competing interests:}
There are no competing interests to declare.
\paragraph*{Data and materials availability:}

Data and materials supporting the findings of this study are available from the corresponding author upon reasonable request. All datasets and code will be deposited in publicly accessible repositories prior to final publication.
\end{acknowledgments}

\newpage
\appendix

\section{Materials and Experimental Methods}

\subsection{Sample preparation}\label{sec:sample}
The sample preparation was performed in ultrahigh vacuum with a base pressure in the low 10$^{-10}$\,mbar range.
After repeated cycles of Ar sputtering ($P_{\mathrm{Ar}}\,=\,1\times10^{-5}$\,mbar; $E\,=\,1$\,keV; 30\,min) and annealing (600$^\circ$C; 30\,min), the Cu(110) substrate was heated to 300$^\circ$C (measured with a DIAS Pyrospot DP 10N pyrometer with $\epsilon=$\,0.05) and exposed to oxygen at 1$\times$10$^{-7}$\,mbar for 10\,min.
This resulted in the well-known p(2$\times$1) oxygen reconstruction, wherein the oxygen atoms are arranged in rows along the [001] direction\,\cite{Kishimoto2008}.
This reconstruction emerges as alternating stripes of bare Cu(110) and (2$\times$1)O-reconstructed Cu(110), where increased exposure leads to wider Cu-O regions\,\cite{Cicoira2006}.
Conversely, overexposure to oxygen leads to the formation of another oxygen-induced reconstruction, namely c(6$\times$2)\,\cite{Kishimoto2008}.
Thus, the exposure time was set as high as possible so that no indications of the c(6$\times$2) structure arose in low-energy electron diffraction (LEED).

Before use in our experiments, the 6T molecules (Tokyo Chemical Industry Co. Ltd.) were purified by sublimation. 
The deposition of the 6T molecules onto the substrate kept at -5$^\circ$C was done using a well-degassed Kentax three-cell evaporator operated at 250$^\circ$C.
The evaporation rate was calibrated with LEED using the findings in Ref.~\cite{BerkebilePhD}.

\subsection{Time-resolved ARPES experiments}\label{sec:trarpes}
The laser setup used for the time-resolved ARPES experiments is based on a Ti:sapphire regenerative amplifier, operated at 200\,kHz, which delivers 800\,nm laser pulses with 40\,fs duration and 8\,$\mu$J pulse energy, which were split into a pump and a probe branch in a 70:30 ratio.

The pulses in the probe branch were further amplified by a two-pass amplifier and subsequently frequency-doubled to 400\,nm, resulting in a pulse duration of about 60\,fs and a pulse energy of 1.5\,$\mu$J. For the generation of high-harmonics, the pulses were tightly focused with an achromatic lens (f = 60\,mm) into a krypton gas jet, which is produced inside a UHV chamber ($10^{-8}$\,mbar without gas load) by injecting pressurized krypton (4\,bar) through a ceramic nozzle with a diameter of 30\,$\mu$m. To separate the 7$^{\textrm{\footnotesize th}}$ harmonic (57\,nm, 21.7\,eV) from the residual harmonics, two wavelength-selective multilayer mirrors were used, yielding an almost isolated 7$^{\textrm{\footnotesize th}}$ harmonic. In addition, the two multilayer mirrors were used to collimate and refocus the p-polarized probe pulses onto the sample (with an angle of incidence of 55$^{\circ}$) after first passing through a 100\,nm thin Al-filter to remove the residual 400\,nm fundamental wavelength.

For optical excitation of the 6T molecules, the pump pulses were frequency-converted to 528\,nm (2.35\,eV) using an optical parametric amplifier (OPA), resulting in a pulse duration of 45\,fs. After passing a linear delay stage --- which delays the pump pulses relative to the probe pulses --- they were focused onto the sample under normal incidence, in order to have the linear polarized pump pulses aligned along the long molecular axis of 6T. However, the resulting non-collinear pump-probe geometry introduces a spatial tilt between both pulse fronts and thus leads to a broadening of the temporal overlap. To compensate for this effect, the pulse fronts of the pump pulses were tilted by focusing the beam with a lens (f=500\,mm) onto an optical grating (f$_{gr}$=1200 g/mm) and subsequently imaging the diffracted spot onto the sample using an additional lens (f=250\,mm).

During measurements the sample was under UHV conditions ($<2\times10^{-10}$\,mbar) and was continuously cooled to T$_{\textrm{s}}=24$\,K with a closed-cycle Helium cryostat for periods of typically 10 days. 
The photoemitted electrons were projected by a momentum microscope --- using a lens system similar to the ones used for photoemission electron microscopy (PEEM) --- onto a time- and position-sensitive delay-line detector after passing though a 1~m drift tube. By measuring for each pump-probe delay time $t$ the time of flight $\tau$ of the photoelectrons as well as the photoelectron intensity $I$ and the position on the detector, which corresponds to the electrons' $k_{x}, k_{y}$ components, we obtain --- after converting the time of flight $\tau$ into a binding energy $E - E_\mathrm{F}$ --- a four-dimensional data set $I(E - E_\mathrm{F},k_{x}, k_{y}, t)$. The energy resolution of our momentum microscope was about 150~meV, and the momentum resolution was better than $0.01$~{\AA}$^{-1}$. Low energy electrons were suppressed by applying a retarding field 
in front of the delay-line detector.

The momentum maps shown in this paper are accumulated for about 25~hours for each delay time. Note that we subtracted a small pump-induced background at the $\bar \Gamma$-Point. During a single delay scan, each data point was accumulated for about 7\,min.

\subsection{Background subtraction for linescans}\label{sec:backgroundsubtraction}

\begin{figure}[h]
	\centering
	\includegraphics[width=1.0\textwidth]{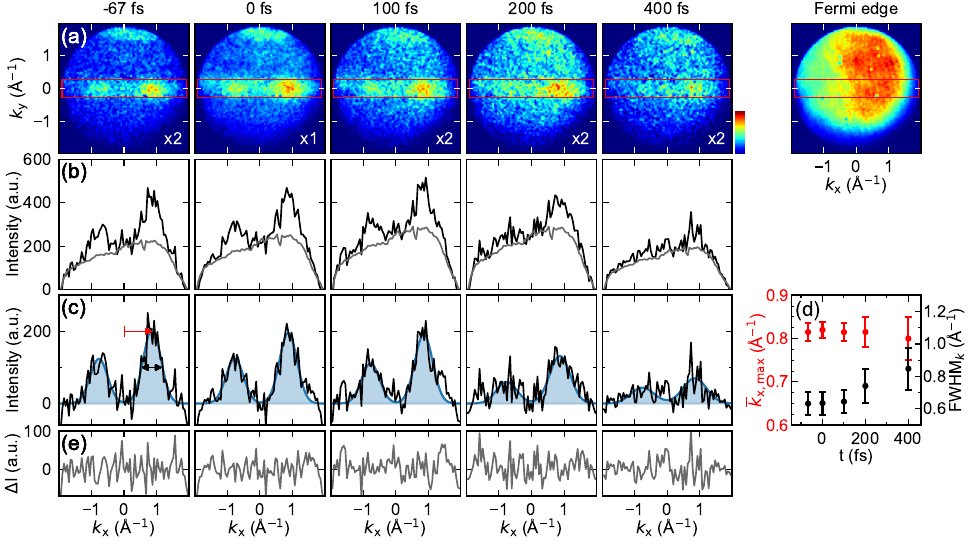}
	\caption{
		(a) Experimental momentum maps at five different pump-probe delays extracted in the energy range of $E - E_\mathrm{F} \in [+0.6, +1.5]$~eV, as well as the momentum map of the Fermi edge taken between $E - E_\mathrm{F} \in [-0.4, +0.4]$~eV.
		(b) Intensity linescans (black) along $k_{x}$ for each delay time extracted within the red boxes in panel (a) together with the fitted linescan of the Fermi edge (gray).
		(c) Background-subtracted linescans (black) fitted by two independent Gaussians with a common $\mathrm{FWHM}_k$ (blue-shaded area).
		(d) Resulting fit parameters of the $\mathrm{FWHM}_k$ and mean peak position as a function of delay time, together with their statistical uncertainties.
		(e) Residues between background- subtracted linescans and fitted Gaussians. 		
	}\label{sifig_backgroundsubtraction}
\end{figure}

In the experiment, the observed exciton signatures appear close to the Fermi edge, which can be seen in the momentum maps in Fig.~\ref{sifig_backgroundsubtraction}a, extracted in the energy range $E - E_\mathrm{F} \in [+0.6, +1.5]$~eV for different delay times.
Therefore, it must be noted that these momentum maps not only contain the exciton signatures, but also a background originating from the Fermi edge. In order to perform a meaningful analysis of the intensity modulation in the exciton momentum maps according to our exciton model, prior subtraction of this background is required.

First, we isolate intensity linescans along $k_{x}$ in the major lobe region in Fig.~\ref{sifig_backgroundsubtraction}a (red boxes, $k_{y}$ = $\pm 0.28${\AA}$^{-1}$) for every delay time. 
Similarly, we obtain an intensity linescan from the momentum map of the Fermi edge, including energies in the range $E - E_\mathrm{F} \in [-0.4, +0.4]$~eV. To ensure a statistically reliable background subtraction, the background caused by the Fermi edge was obtained from the same data cube (but in the lower energy range $[-0.4, +0.4]$~eV) rather than from a measurement recorded in the same energy range $[+0.6, +1.5]$~eV as the signal, but without the pump pulse.
This approach was chosen because the data acquired without the pump pulse exhibit insufficient statistics of the count rate in the momentum map, making them unsuitable for accurate background determination. In contrast, since the momentum distribution of the Fermi edge is nearly identical in both energy regions --- apart from variations in the intensity --- using the $[-0.4, +0.4]$~eV data provides improved statistics and thus a more robust background subtraction for our analysis. 

The extracted linescans within the red boxes (black) and the fitted Fermi background (gray) are plotted in panel (b) of Fig.~\ref{sifig_backgroundsubtraction}. The same Fermi background was fitted to each of the linescans using two parameters - a scaling factor and an offset. The resulting background-subtracted linescans are displayed in Fig.~\ref{sifig_backgroundsubtraction}c together with a fit function (blue-shaded area). According to our exciton model, the fit function consists of two Gaussians with five adjustable parameters: two peak positions $k_{x, \text{max}}^{-}$ and $k_{x, \text{max}}^{+}$ and two amplitudes (to allow for a small misalignment of our sample), and one common $\mathrm{FWHM}_k$. The obtained fit parameters and their statistical uncertainties are presented in Fig.~\ref{sifig_backgroundsubtraction}d. Note that $\bar{k}_{x, \text{max}} = \frac{1}{2} (|k_{x, \text{max}}^{-}| + k_{x, \text{max}}^{+})$ here corresponds to the mean value of the two fitted peak positions. The residuals between background-subtracted linescans and the fit function are displayed in Fig.~\ref{sifig_backgroundsubtraction}e. Further analysis of the data, as well as a compressed version of Fig.~\ref{sifig_backgroundsubtraction} is provided in the main text (see Fig.~\ref{fig4}). Note, however, that in case of Fig.~\ref{fig4} the Fermi edge has already been subtracted from the displayed exciton momentum maps. For this, we used the same parameters (scaling factor, offset) as for the background subtraction in the linescans.

\subsection{Temporal evolution of exciton energy}\label{sec:energyspectrum}

\begin{figure}[h]
	\centering
	\includegraphics[width=1.0\textwidth]{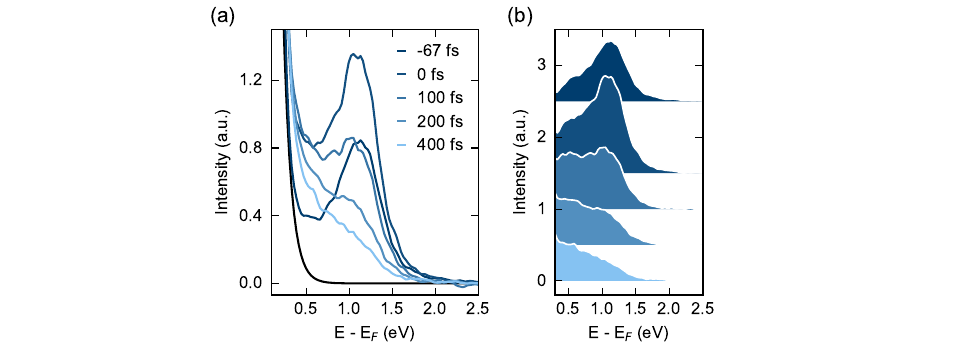}
	\caption{
		(a) Energy spectra extracted within the $k$ region defined by the red boxes in Fig.~\ref{sifig_backgroundsubtraction}a, showing the exciton signal at five different pump-probe delays, along with the Fermi edge (black) derived from a data cube measured without the pump pulse.
		(b) Remaining energy spectra of the exciton after subtraction of the Fermi edge.
	}\label{sifig_energyspectrum}
\end{figure}

The analysis of background-subtracted linescans, performed within the framework of our exciton model, reveals a temporal increase in the $\mathrm{FWHM}_k$ from $t=0$~fs to $t=400$~fs, implying a progressive localization of the exciton wave function. This reduction of spatial extent should likewise be observable in the energy spectrum as an increase of the exciton binding energy and a concomitant decrease of the excited-state energy, over time. Note that a decrease of the excited-state energy also corresponds to an energy shift towards lower final-state (kinetic) energies in a two-photon photoemission experiment. 
In Fig.~\ref{sifig_energyspectrum}a, we therefore plot for different delay times the energy spectrum extracted from the same $k$ region (red boxes) as depicted in Fig.~\ref{sifig_backgroundsubtraction}a. To identify a possible shift of the excited-state energy, we first subtract the Fermi edge (black curve) for all delay times, resulting in the background-subtracted energy spectra in Fig.~\ref{sifig_energyspectrum}b. Here, the energy spectrum of the Fermi edge was obtained from a data cube measured without pump pulse, using the same $k$ region (red boxes) as for the exciton signatures. A closer inspection of the energetic position of the exciton in Fig.~\ref{sifig_energyspectrum}b indeed reveals a shift towards lower final-state energies for longer delay times. We attribute this energy shift to two separate contributions: The first effect is caused by the optical-excitation of the exciton being slightly above resonance, leading to an ultrafast, small energy relaxation at early delay times. This process is also observed in the momentum-integrated intensity of the exciton as a function of delay time in Fig.~\ref{fig2}c, where for early delay times ($t=0-100$~fs) a fast-decaying component due to an off-resonant excitation dominates the signal, whereas at later delay times ($t=100-1000$~fs) the slower component due to the transiently populated exciton takes over. The energy shift for $t>100$~fs we attribute to a second effect, namely an increase of the exciton binding energy accompanied by a reduction of its spatial extent over time, which ultimately supports our conclusions from the background-subtracted linescan analysis.

\section{Computational Methods}\label{sec:compmethod}

\subsection{\emph{Ab initio} calculations for clusters}\label{sec:cluster}

\begin{figure}[h]
\centering
\includegraphics[width=0.6\textwidth]{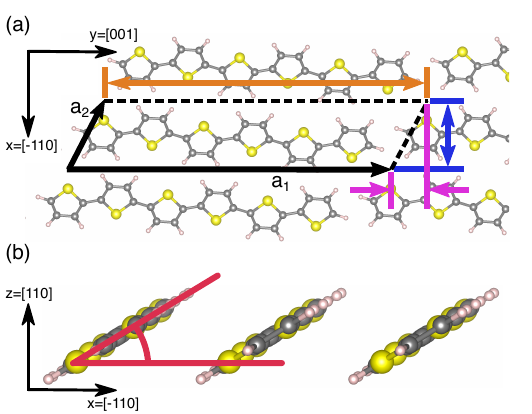}
\caption{(a) Top-down view of the monolayer: The unit cell vectors are shown in black, where a$_2$ determines the row-slip (purple) of 2.807~{\AA}. The row-distance  $d$ (blue) equals 5.335~{\AA}, and the unit cell length (orange) is 25.258~{\AA}. (b) Side view of the monolayer: The molecules are tilted (red) by 31.5$^{\circ}$ relative to the $xy$-plane. }\label{sifig1}
\end{figure}

In this section, we provide details of our GW/BSE calculations for an isolated 6T molecule and a cluster containing four molecules. The geometries of these two systems were obtained by cutting out one/four molecules along the $a_2$ direction from the periodic monolayer shown in Fig.~\ref{sifig1}.

In a first step, we performed single-point ground-state density functional theory (DFT) calculations using ORCA 5.0.4 \cite{ORCA, ORCA5} with the 6-311G* Gaussian-orbital basis set \cite{krishnan1980self, mclean1980contracted} and the B3LYP hybrid functional \cite{becke1993density, stephens1994ab}. 
The resulting Kohn-Sham orbitals and energies then served as the basis for the GW-calculation. For this second step, we employed the Fiesta code \cite{jacquemin2017bethe}. Specifically, we performed an iterative, self-consistent correction of the Kohn-Sham energies while keeping the wave functions unchanged.
For a single 6T molecule, we explicitly corrected 10 occupied (valence) and 10 unoccupied (conduction) orbitals. Orbitals beyond there were rigidly shifted by the correction that was applied to the HOMO-9/LUMO+9. For the tetramer calculation, we explicitly corrected 40 occupied and 40 unoccupied orbitals.

In the third step, we solved the Bethe-Salpeter equation (BSE) in which excited states are described as correlated electron-hole pairs \cite{Rohlfing2000a, blase2020bethe}
\begin{equation}
\label{eq:BSEcluster}
\sum_{v'c'} H_{vc,v'c'} X_{v'c'}^{m} = \Omega_{m} X_{vc}^{m}.
\end{equation}
Here, the eigenvalues $\Omega_{m}$ of the effective electron-hole Hamiltonian are the system's charge neutral excitation energies and the eigenvectors $X_{vc}^{m}$ encode the entangled nature of the electron-hole state, which can be expressed via the exciton wave function as
\begin{equation}
\label{eq:Psicluster}
\Psi_{m}(\ve{r}_e, \ve{r}_h) = \sum_{vc} X_{vc}^{m} \varphi_{v}^* (\ve{r}_h) \xi_{c}(\ve{r}_e).
\end{equation}
Using again the Fiesta code, we iteratively diagonalized the BSE Hamiltonian and obtained the first 20 excited states of the single molecule and the tetramer. In both cases, we constructed the BSE Hamiltonian by including all valence orbitals up to 20~eV below the HOMO and all conduction orbitals up to 20~eV above the LUMO. This resulted in the use of 72 valence and 135 conduction orbitals in the single-molecule calculation and 291 valence and 507 conduction orbitals in the tetramer calculation. In both cases, we used the Tamm-Dancoff approximation (TDA), since test calculations showed de-excitations to contribute less than 1\% to the BSE eigenvector. 

In the final step, we simulated photoemission momentum maps from excitons using the exPOT approach described previously \cite{Kern2023}. Denoting the energy of the probe laser as $\omega$ and the kinetic energy of the photoelectron as $E_\mathrm{kin}$, the intensity of photoemission from exciton $m$ as a function of the photoelectron momentum $\ve k$ is given by  \cite{Kern2023}
\begin{equation}
\label{eq:exPOTcluster}
I_{m}(\ve{k}) \propto
\sum_v 
\left| \sum_c X_{vc}^{m} \mathcal{F} \left[ \xi_{c} \right](\ve{k}) \right|^2 \cdot
\delta\left(\omega + \Omega_{m} - \varepsilon_{v} - E_\mathrm{kin}  \right).
\end{equation}
The formula shows that the hole state enters the energy conservation term via its single-particle energy $\varepsilon_{v}$, while the angular distribution is determined by a coherent sum of the Fourier transforms $\mathcal{F}$ of the electron orbitals contributing to the exciton with weights $ X_{vc}^{m}$. 
Note, that in Eq.~\ref{eq:exPOTcluster} the polarization factor $\left|\ve A \cdot \ve k\right|^2$ has been omitted.

\subsection{\emph{Ab initio} calculations for a periodic 6T layer}\label{sec:periodic}

This section summarizes details of our \emph{ab initio} GW/BSE calculations  for a free-standing, periodic monolayer of 6T. Its geometry is based on the experimentally determined structure of a 6T crystal \cite{Horowitz1995}. We use a skewed unit cell containing one molecule per cell as depicted in Fig.~\ref{sifig1}.
Just as in the cluster case, we perform a series of DFT, GW and BSE calculations to get all the quantities required by the exPOT formalism, Eq.~\ref{eq:exPOTperiodic}.

For the first step, the ground-state DFT calculations, we used Quantum Espresso (PWSCF v.7.2) \cite{Giannozzi2009, Giannozzi2017, Giannozzi2020}, employing the PBE functional \cite{perdew1996generalized} and a plane-wave energy cutoff of 70~Ry for wave functions and 280~Ry for the density. To avoid spurious interactions between repeated slabs in the $z$ direction, the unit cell height was set to 28.525~{\AA}, which corresponds to a vacuum layer of 25.521~{\AA}. Additionally, we employed a truncation scheme for the Coulomb interaction \cite{Sohier2017}.
For this and all upcoming steps, unless stated otherwise, we utilized a $2\times8\times1$ $k$-mesh.

Steps two and three, the GW and BSE calculations, were performed using BerkeleyGW~4.0 \cite{Hybertsen1986, Rohlfing2000a, Deslippe2012}.
Before performing the GW step, we employed BerkeleyGW's ParaBands tool to calculate stochastic conduction pseudobands \cite{altman2024mixed} to reduce calculation times and scaling of the subsequent GW-BSE calculation. We protected 27 conduction bands, which were then utilized for the construction of the BSE Hamiltonian. The remaining conduction bands were compressed into 520 pseudobands, resulting in a total of 620 bands (including the 73 valence bands).
The actual GW calculation was then conducted at the $G_0 W_0$ level. For faster convergence of the $G_0 W_0$ energies with respect to the $k$-mesh, we used the NNS subsampling technique \cite{da_Jornada2017}. 
As a substep, we calculated the static inverse dielectric function by employing a cutoff energy of 20~Ry for reciprocal lattice vectors and summing over 500 bands.
Finally, we obtained the $G_0W_0$ energies of 27 valence and 27 conduction bands by summing again over 500 bands and using the same plane-wave cutoff as before.
Then, for use in the ensuing BSE calculation, we recalculated the static inverse dielectric function applying the same settings as above, but without the NNS subsampling. 

For solving the BSE, we considered 27 valence and 27 conduction bands in the construction of the direct and exchange kernel matrix.
Note that for solving the BSE, we calculated wave functions on a finer $k$-mesh, $10\times40\times1$, and interpolated the $G_0W_0$ corrections and the direct and exchange kernel to the finer grid (interpolating from the 27 coarse valence and 27 coarse conduction wave functions to 20 fine valence and 22 fine conduction wave functions). Then, by fully diagonalizing the BSE Hamiltonian, and employing the Tamm-Dancoff approximation, 
\begin{equation}
\label{eq:BSEperiodic}
\sum_{v'c'\ve{q}'} H_{vc\ve{q},v'c'\ve{q}'} X_{v'c'\ve{q}'}^{m} = \Omega_{m} X_{vc\ve{q}}^{m},
\end{equation}
we obtained the exciton energies and eigenvectors for optical excitations, that is, excitations with center-of-mass momentum $\ve Q=0$. 
As before, we utilized the BSE eigenvectors to construct the electron-hole wave functions
\begin{equation}
\label{eq:Psiperiodic}
\Psi_{m}(\ve{r}_e, \ve{r}_h) = \sum_{vc} \sum_{\ve{q}}^{\mathrm{BZ}} X_{vc\ve{q}}^{m} \varphi_{v\ve{q}}^* (\ve{r}_h) \xi_{c\ve{q}}(\ve{r}_e).
\end{equation}

In the final step, we simulated photoemission angular distributions also for excitons in the periodic 6T layer, using the recently described exPOT formalism for periodic systems \cite{Kaidisch2025}. For zero center-of-mass momentum and omitting the $\left|\ve A \cdot \ve k\right|^2$ polarization factor, the photoemission intensity becomes
\begin{equation}
\label{eq:exPOTperiodic}
I_{m}(\ve{k}) \propto
\sum_v \sum_{\ve{q}}^{\mathrm{BZ}}
\left| \sum_c X_{vc\ve{q}}^{m} \mathcal{F} \left[ \xi_{c\ve{q}} \right](\ve{k}) \right|^2 \cdot
\delta\left(\omega + \Omega_{m} - \varepsilon_{v\ve{q}} - E_\mathrm{kin} \right).
\end{equation}
Note the additional sum over the Brillouin zone and the dependence on the Bloch vector $\ve{q}$ when comparing to the corresponding Eq.~\ref{eq:exPOTcluster} for non-periodic clusters.

\begin{figure}[h]
\centering
\includegraphics[width=0.8\textwidth]{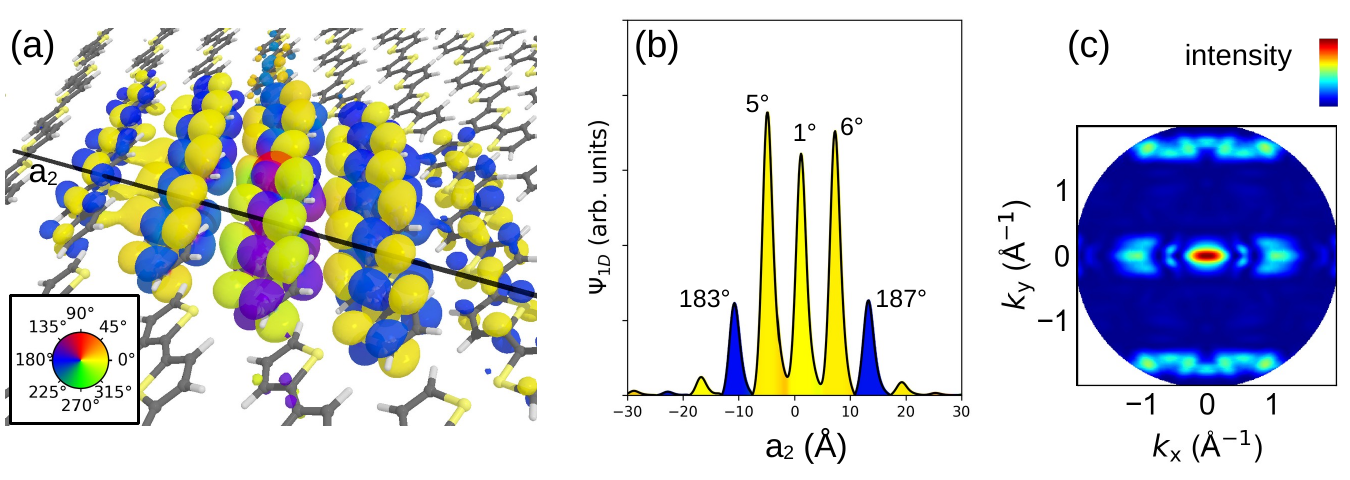}
\caption{ 
In analogy to Fig.~\ref{fig3}: (a) Fixed-hole exciton wave function. (b) Line cut through the wave function along the $a_2$ line. (c) Momentum map, calculated with Eq.~\ref{eq:exPOTperiodicsimple}, using the BSE eigenvector of $S_3$ instead of $S_1$.
}\label{sifig2}
\end{figure}

The so-obtained exciton wave function and momentum map for $S_1$ are shown in Fig.~\ref{fig3}b--d. An analogous visualization of $S_3$ is given in Fig.~\ref{sifig2}.
In both cases, the exciton wave function takes approximately the form of a combination of isolated molecule LUMOs, located at different molecular sites (Fig.~\ref{sifig2}a). A line cut of the exciton wave function exhibits maxima at each molecular site (Fig.~\ref{sifig2}b).
Fitting a Gaussian envelope function to these maxima, we extract a $\mathrm{FWHM}_x$ of $\sim 16.0$~{\AA} for $S_1$, while we find $\sim 19.9$~{\AA} to be fitting for $S_3$. 
Note, that while these values serve as a rough estimate for the exciton delocalization, they should be taken with a grain of salt, since both $S_1$ and, especially, $S_3$ don't follow a perfect Gaussian, with in particular the central maximum being too low.
It is noteworthy that the change of the complex phase between neighboring molecules is quite distinct between the two excitons: for $S_1$, we find a constant change of $\sim 195^\circ$ between neighboring molecules, while for $S_3$, the phase either changes by $\sim 180^\circ$ or $\sim 0^\circ$. 
Fig.~\ref{sifig4} (c) and (d) depict simulated momentum maps for $S_1$ and $S_3$. Their comparison clearly highlights the role of the phase in modulating the intensity in the momentum maps. 

\subsection{Comparison of cluster and periodic \emph{ab initio} results}\label{sec:compresults}

\begin{figure}[h]
\centering
\includegraphics[width=0.6\textwidth]{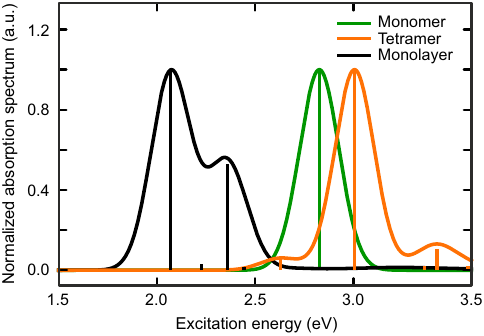}
\caption{The spectrum of the monomer is dominated by a single optically active exciton at 2.826~eV, which is also the lowest-energy excitation of the system. For the tetramer, the most active exciton is the fourth one, located at 3.005~eV. In the monolayer system, the lowest-energy exciton is located at 2.070~eV and is the most optically active one, followed by the third exciton which is located at 2.358~eV. All spectra have been normalized individually to a maximum value of 1.0.}\label{sifig3}
\end{figure}

In this section we compare key properties of the three systems for which we have performed GW/BSE calculations: (i) the monomer, (ii) the tetramer, and (iii) the periodic free-standing layer. The calculated optical absorption spectra for light polarized along the long molecular axis ($y$ direction) are shown in Fig.~\ref{sifig3}.  Note that each spectrum was normalized to a maximum value of 1.0 for better comparability.
The monomer spectrum shows a single optically active exciton, while for the tetramer additional, albeit much weaker, optically active excitons appear.
For the monolayer, we observe an overall redshift of the spectrum, which is due to the different computational treatment of the monomer and tetramer as opposed to the extended monolayer. For the latter, we used a single-shot $G_0W_0$ calculation on top of a PBE starting point, while for the former we could afford B3LYP starting points and self-consistent $GW$ corrections. The spectrum of the free-standing monolayer is dominated by two optically active excitons, where the first exciton, $S_1$, exhibits a transition dipole moment about twice as large as the one of the next optically active exciton, $S_3$.

The compositions of these optically active excitons are summarized in Table~\ref{sitable1}, where the transition probability from valence orbital $v$ to conduction orbital $c$ in exciton $m$ is given by 
\begin{equation}
    p^{m}_{v \rightarrow c} = \left|  X_{vc}^{m} \right|^2
\end{equation}
in the cluster case and
\begin{equation}
    p^{m}_{v \rightarrow c} = \sum_{\ve{q}}^{\mathrm{BZ}} \left|  X_{vc\ve{q}}^{m} \right|^2
\end{equation}
for the periodic monolayer. Specifically, we find that the contribution of the HOMO-1 to LUMO+1 transition shrinks from 8.4\% in the monomer case, to 6.2\% in the tetramer and only $2-3\%$ in the monolayer system. Conversely, going from the clusters to the monolayer, the HOMO to LUMO transition grows even more in importance, fully dominating all other contributions to the excitons. 
As a technical note regarding this analysis we point out that, in the case of the tetramer, there are four orbitals with monomer-HOMO character and four orbitals with monomer-LUMO character, and so on. Thus, when calculating the contribution of a particular transition to a given exciton, e.g., the HOMO to LUMO transition of the tetramer, we sum $p^{m}_{v \rightarrow c}$ over these four valence and conduction orbitals (making it a sum of 16 contributions). 

\begin{table}[h]
\centering
\caption{Properties of dominant optically active excitons of the monomer, tetramer, and monolayer: For each system, the transition dipole moment in the direction of the light polarization $|d_y|^2$ of the optically strongest exciton is normalized to 1. The last four columns contain the transition probabilities from HOMO/HOMO-1 to LUMO/LUMO+1 for each of the excitons.}\label{sitable1}
\begin{tabular}{lccccccc}
\hline
System & Exciton & Excitation Energy (eV) & $|d_y|^2$ & H$\rightarrow$L& H-1$\rightarrow$L+1 & H$\rightarrow$L+1 & H-1$\rightarrow$L \\\\
\hline
Monomer & $S_1$ & 2.826 & 1.000 & 0.884& 0.084 & 0.000 & 0.000  \\ \\
Tetramer & $S_4$ & 3.005 & 1.000 & 0.883& 0.062 & 0.008 & 0.010  \\
\\
Monolayer & $S_1$ & 2.070 & 1.000 & 0.931& 0.031 & 0.006 & 0.008  \\
Monolayer & $S_3$ & 2.358 & 0.529 & 0.916& 0.021 & 0.025 & 0.017  \\
\hline
\end{tabular}
\end{table}

\begin{figure}[h]
\centering
\includegraphics[width=1.0\textwidth]{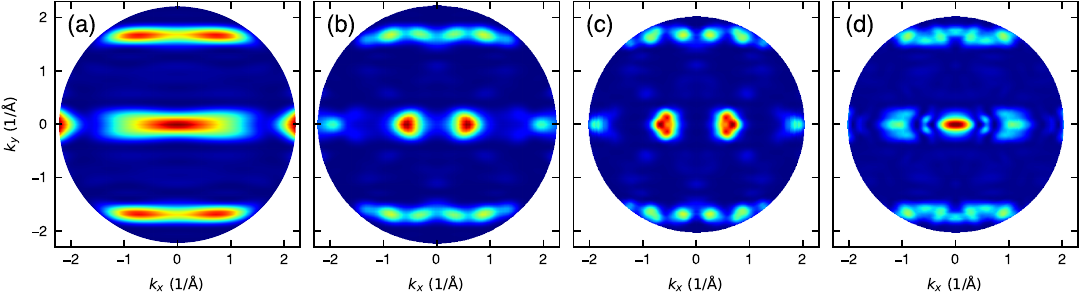}
\caption{
(a) Simulated momentum map for $S_1$ of the monomer, (b) $S_4$ of the tetramer, (c) $S_1$ of the monolayer and (d) $S_3$ of the monolayer. Panels (a)--(c) show emission from the dominant optically bright excitons of the three systems and show a clear progression towards an increase in the separation between the two intensity peaks along $k_y=0$. In contrast, the optically less bright $S_3$ of the monolayer system, panel (d), exhibits a strong maximum around the $\bar\Gamma$-point.}\label{sifig4}
\end{figure}

Finally, we use the exPOT formalism to calculate photoemission momentum maps for all systems listed in Table~\ref{sitable1}. 
The results are shown in Fig.~\ref{sifig4}. A few points are noteworthy. 
First, the monomer does not exhibit the intensity modulation along the $k_y=0$ line.
Secondly, the intensity modulation already appears for the tetramer, however, with a smaller separation of the intensity peaks, which are located at $k_{x,\max} \approx \pm 0.57$~{\AA}$^{-1}$, when compared to the experimental positions of $k_{x,\max} \approx \pm 0.85$~{\AA}$^{-1}$. This value increases to $k_{x,\max} \approx \pm 0.64$~{\AA}$^{-1}$ in the case of $S_1$ of the monolayer system.
Finally, it shall be noted, that the optically bright $S_3$ exciton of the monolayer also exhibits an intensity modulation along the $k_y=0$ line. 
However, owing to the different phase modulation of the $S_3$ exciton wave function compared to the one of $S_1$, the intensity peak appears at the $\bar \Gamma$ point. 
Moreover, for $S_1$ and $S_3$ we observe slightly different intensity distributions in the minor lobe region at $k_y\approx \pm 1.7$~{\AA}$^{-1}$.
However, because of the only $\sim 0.3$~eV higher excitation energy of $S_3$ and its weaker transition dipole moment compared to $S_1$ (see Fig.~\ref{sifig3}), the present time-resolved POT experiments are not able to reveal the presence of $S_3$.

\section{Relation of exciton model with first-principles description}\label{sec:modelvsabinitio}

\subsection{exPOT formalism with wannierized Dyson orbitals}\label{subsec:wannierDyson}

The aim of this section is to show that the equation for the photoemission intensity from an excited state in a periodic system, derived in the plane-wave approximation and given in Ref.~\cite{Kaidisch2025}, can, in its most general form, be compactly written via localized Dyson orbitals, in close analogy to ground-state, non-periodic POT (see, e.g., Eq.~10 in Ref.~\cite{Dauth2014}). 

To show this, we start from the general photoemission expression given in Ref.~\cite{Kaidisch2025}
\begin{align}
  \label{eq:expotperiodic_pw_version}
  I_{m\ve Q}(\ve k) \propto &|\ve A \cdot \ve k|^2 \sum_v 
  \left|\sum_{c} X_{vc\ve{q}}^{m \ve Q} \mathcal F [\xi_{c\ve{q}+\ve Q}](\ve q + \ve G + \ve Q)\right|^2 
   \cdot \delta\left(\omega + \Omega_{m \ve Q} -\varepsilon_{v\ve{q}} - E_\mathrm{kin} \right), \nonumber \\
  & \mathrm{with}~\ve q + \ve G + \ve Q= \ve k,
\end{align}
which, in contrast to Eq.~\ref{eq:exPOTperiodic}, also allows for non-zero center-of-mass momentum $\ve Q$. Moreover, it explicitly includes the polarization factor, thus keeping the derivation as general as possible.
Also note that in this formulation, the sum over $\ve q$ has been eliminated as explained in more detail in Ref.~\cite{Kaidisch2025}. Briefly, we use the fact that
\begin{align}
\label{eq:ft_of_bloch}
\mathcal F [\xi_{c \ve{q}+\ve Q}](\ve k) = \int_V \mathrm{d}^3 r \xi_{c \ve{q}+\ve Q}(\ve r) e^{-\mathrm{i}\ve k \cdot \ve r}
\end{align}
vanishes, unless $\ve q + \ve G + \ve Q = \ve k $ for at least one of the reciprocal lattice vectors $\ve G$. And since, for an arbitrary $\ve k$, this condition is either not fulfilled at all, or, at best, satisfied by a single $\ve q$ (and $\ve G$), we can drop the sum and restrict $\ve k$ to the $\ve q + \ve G + \ve Q$ grid spanned by the $\ve q$ and $\ve G$ vectors.

Using the same argument, we can write
\begin{align}
  \mathcal F [\xi_{c \ve{q}+\ve Q}](\ve q + \ve G + \ve Q) = \sum_{\ve{q'}}^{\mathrm{BZ}} \mathcal F [\xi_{c \ve{q'}+\ve Q}](\ve q + \ve G + \ve Q).
\end{align}
Inserting this into Eq.~\ref{eq:expotperiodic_pw_version} and using the linearity of the Fourier transform, we find
\begin{align}
  \label{eq:expotperiodic_pw_version2}
  I_{m \ve Q}(\ve k) \propto &|\ve A \cdot \ve k|^2 \sum_v
  \left|\mathcal F \left[\sum_{\ve{q'}}^{\mathrm{BZ}} \sum_{c} X_{vc\ve{q'}}^{m \ve Q} \xi_{c \ve{q'}+\ve Q}\right](\ve q + \ve G + \ve Q)\right|^2 \cdot
   \delta\left(\omega + \Omega_{m \ve Q} -\varepsilon_{v \ve{q}} - E_\mathrm{kin} \right), \nonumber \\
  & \mathrm{with}~\ve q + \ve G + \ve Q= \ve k.
\end{align}
Next, we use the Dyson orbitals as defined in Ref.~\cite{Kaidisch2025} as
\begin{align}
  D_{v \ve{q}}^{m \ve Q}(\ve r) = \sum_{c}
  X_{v c \ve{q}}^{m \ve Q}
  \xi_{c \ve{q}+\ve Q}(\ve r)
\end{align}
to observe that
\begin{align}
  \sum_{\ve{q'}}^{\mathrm{BZ}} \sum_{c} X_{v c \ve{q'}}^{m \ve Q} \xi_{c \ve{q'}+\ve Q}(\ve r) 
  &=\sum_{\ve{q'}}^{\mathrm{BZ}} D_{v \ve{q'}}^{m \ve Q}(\ve r),
\end{align}
which we insert into Eq.~\ref{eq:expotperiodic_pw_version2} to get
\begin{align}
  \label{eq:expotperiodic_pw_version3}
  I_{m \ve Q}(\ve k) \propto &|\ve A \cdot \ve k|^2 \sum_v 
  \left|\mathcal F \left[\sum_{\ve{q'}}^{\mathrm{BZ}} D_{v \ve{q'}}^{m \ve Q}\right](\ve q + \ve G + \ve Q)\right|^2 \cdot
   \delta\left(\omega + \Omega_{m \ve Q} -\varepsilon_{v \ve{q}} - E_\mathrm{kin} \right), \nonumber \\
  & \mathrm{with}~\ve q + \ve G + \ve Q= \ve k.
\end{align}
Finally, since in the Brillouin-zone summation only a single $\ve q'$ contributes and since $\left|e^{-\mathrm i (\ve q' + \ve Q) \cdot \ve R}\right|^2=1$ (where $\ve R$ is a lattice vector), we can write
\begin{align}
  \label{eq:expotperiodic_pw_version4}
  I_{m \ve Q}(\ve k) \propto &|\ve A \cdot \ve k|^2 \sum_v 
  \left|\mathcal F \left[\frac{1}{\sqrt{N_{q'}}}\sum_{\ve{q'}}^{\mathrm{BZ}} e^{-\mathrm i (\ve q' + \ve Q) \cdot \ve R}D_{v\ve{q'}}^{m \ve Q}\right](\ve q + \ve G + \ve Q)\right|^2 \cdot
   \delta\left(\omega + \Omega_{m \ve Q} -\varepsilon_{v \ve{q}} - E_\mathrm{kin} \right), \nonumber \\
  & \mathrm{with}~\ve q + \ve G + \ve Q= \ve k,
\end{align}
where $N_{q'}=N_q$ is the number of $\ve q$ vectors in the used mesh, i.e., the number of unit cells in the crystal, employing as usual Born-von Kármán periodic boundary conditions.

The expression inside the Fourier transform can be identified as a wannierized Dyson orbital $D_{v \ve{R}}^{m \ve Q}$ at lattice site $\ve{R}$ \cite{wannier1937structure}
\begin{align}
	\label{eq:wannier_dyson}
	D_{v \ve{R}}^{m \ve Q}(\ve r)
	&= \frac{1}{\sqrt{N_{q'}}}
	\sum_{\ve{q'}}^{\mathrm{BZ}} e^{-\mathrm i (\ve q' + \ve Q) \cdot \ve R}D_{v\ve{q'}}^{m \ve Q}(\ve r) \nonumber \\ 
	&= \frac{1}{\sqrt{N_{q'}}}
	\sum_{\ve{q'}}^{\mathrm{BZ}} \sum_{c} 
	e^{-\mathrm i (\ve q' + \ve Q) \cdot \ve R}
	X_{v c \ve{q'}}^{m \ve Q} \xi_{c \ve{q'}+\ve Q}(\ve r),
\end{align}
allowing us to write 
\begin{align}
  \label{eq:expotperiodic_pw_version5}
  I_{m \ve Q}(\ve k) \propto &|\ve A \cdot \ve k|^2 \sum_v 
  \left|\mathcal F \left[D_{v \ve{R}}^{m \ve Q}\right](\ve q + \ve G + \ve Q)\right|^2 \cdot
   \delta\left(\omega + \Omega_{m \ve Q} -\varepsilon_{v \ve{q}} - E_\mathrm{kin} \right), \nonumber \\
  & \mathrm{with}~\ve q + \ve G + \ve Q= \ve k.
\end{align}
Based on this final expression, we can interpret photoemission from an excited state as resulting from a sum over the valence bands of the system, where each band contributes via the Fourier-transform of a Wannier-Dyson orbital (built from conduction band wave functions), modulated by a band-structure-dependent energy-conserving function.
The advantage of this particular formulation is that the use of a Wannier-Dyson orbital naturally reflects the localized character of the excited state, which makes the further analysis  of the latter particularly lucid, similar to a recently proposed scheme \cite{Tao2025}.
We also note that the photoemission expression Eq.~\ref{eq:expotperiodic_pw_version5} is independent of the position $\ve R$ of the Wannier-Dyson orbital, reflecting the periodic nature of the crystal.

Having this new formulation at hand, we can now apply it to the special case of the 6T monolayer excitons $S_1$ and $S_3$ discussed above. These excitons have a center of mass momentum $\ve Q = 0$ and are, to a good approximation, comprised of only transitions between the HOMO band and LUMO band (see Table~\ref{sitable1}).
Dropping the polarization factor for simplicity and choosing $\ve R=0$ (since the position of the Wannier-Dyson orbital does not matter, as shown above), Eq.~\ref{eq:expotperiodic_pw_version5} becomes
\begin{align}
  \label{eq:expotperiodic_pw_version6}
  I_{m}(\ve k) \propto &
  \left|\mathcal F \left[D_{v_1 \ve{R}=0}^{m}\right](\ve{q} +\ve G)\right|^2 \cdot
  \delta\left(\omega + \Omega_{m} -\varepsilon_{v_1 \ve{q}} - E_\mathrm{kin} \right) 
  , \nonumber \\
  & \mathrm{with}~\ve q + \ve G = \ve k.
\end{align}
Additionally, in analogy to the limited energy resolution in the experiments, we integrate Eq.~\ref{eq:expotperiodic_pw_version6} over the kinetic energy.
To calculate a kinetic-energy integrated momentum map, we integrate over $k_z$ at each position $(k_x, k_y)$ of the so-obtained map
\begin{align}
  \label{eq:expotperiodic_pw_version7}
  I_{m}(k_x, k_y) \propto &
  \int_{0}^{\infty} d_{k_z}
  \left|\mathcal F \left[D_{v_1 \ve{R}=0}^{m}\right](\ve{q} +\ve G)\right|^2 \cdot
  \delta\left(\omega + \Omega_{m} -\varepsilon_{v_1 \ve{q}} - \frac{1}{2}(k_x^2+k_y^2+k_z^2) \right) 
  , \nonumber \\
  & \mathrm{with}~\ve q + \ve G = \ve k.
\end{align}
Since $q_z=0$ (as usual, our repeated slab calculation does not sample the $q_z$ direction) and $q_x$ and $q_y$ are fixed at each $(k_x, k_y)$ by $\ve q + \ve G = \ve k$, inside the integral for a given $(k_x, k_y)$ we find $\ve q$ to be a constant vector and $\varepsilon_{v_1 \ve{q}}$ a constant number. 
Therefore also $\omega + \Omega_{m} -\varepsilon_{v_1 \ve{q}}$ is a constant and the delta function is satisfied only for a single value of $k_z$, namely
\begin{equation}
	\label{eq:k_z}
	k_z = \sqrt{ 2 \left(\omega + \Omega_{m} -\varepsilon_{v_1 \ve{q}}\right) - k_x^2 - k_y^2}.
\end{equation}
Thus, we finally obtain the kinetic-energy integrated photoemission expression
\begin{align}
  \label{eq:expotperiodic_pw_version8}
  I_{m}(k_x, k_y) \propto &
  \left|\mathcal F \left[D_{v_1 \ve{R}=0}^{m}\right](\ve{q} +\ve G)\right|^2
  , \nonumber \\
  & \mathrm{with}~ q_{x} + G_{x} = k_{x},~ q_{y} + G_{y} = k_{y}
  \nonumber \\
  & \mathrm{and}~ q_{z} = 0,~ G_{z} = \sqrt{ 2 \left(\omega + \Omega_{m} -\varepsilon_{v_1 \ve{q}}\right) - k_x^2 - k_y^2}.
\end{align}
Note, that in the case of a flat HOMO band, $\varepsilon_{v_1 \ve{q}} = \varepsilon_{v_1}$, this trivially equates to simply using Eq.~\ref{eq:expotperiodic_pw_version6} at the $\ve k$ vectors satisfying $E_\mathrm{kin} = \frac{1}{2}|\ve k|^2 = \omega + \Omega_{m} -\varepsilon_{v_1}$, with the delta function being effectively set to $1$.
Given a high-enough unit cell in the calculation, the spacing of reciprocal lattice vectors $\ve G$ in $k_z$ direction becomes fine enough, such that the condition on $G_z$ can be well met.
To avoid overly lengthy formulas during the rest of this section, we compactly write Eq.~\ref{eq:expotperiodic_pw_version8} as
\begin{align}
  \label{eq:expotperiodic_6T1}
  I_{m}(\ve k) \propto &
  \left|\mathcal F \left[D_{v_1 \ve{R}=0}^{m}\right](\ve{q} +\ve G)\right|^2, \nonumber \\
  & \mathrm{with}~\ve q + \ve G = \ve k.
\end{align}
Note therefore, that whenever Eq.~\ref{eq:expotperiodic_6T1} is referenced, Eq.~\ref{eq:expotperiodic_pw_version8} represents its formally correct counterpart as the kinetic-energy integrated photoemission intensity. 
Similarly, all upcoming equations derived from Eq.~\ref{eq:expotperiodic_6T1} are to be understood as compact versions of their formally correct counterparts analogous to Eq.~\ref{eq:expotperiodic_pw_version8}.

Based on Eq.~\ref{eq:expotperiodic_6T1}, the photoemission intensity is simply the absolute square of a single Fourier-transformed Dyson orbital, which itself is, according to Eq.~\ref{eq:wannier_dyson}, composed of LUMO band Bloch functions
\begin{align}
  \label{eq:wannier_dyson2}
  &D_{v_1 \ve{R}=0}^{m}(\ve r) =
  \frac{1}{\sqrt{N_{q}}}\sum_{\ve{q}}^{\mathrm{BZ}} X_{v_1 c_1 \ve{q}}^{m} \xi_{c_1\ve{q}}(\ve r).
\end{align}

\subsection{Derivation of the exciton model as wannierized Dyson orbital}\label{subsec:derivemodel}

By creating a model that emulates the Dyson orbital, Eq.~\ref{eq:wannier_dyson2}, we can directly simulate the photoemission process based on an easy-to-understand model function, without the need for expensive calculations.
To arrive at the model Eq.~\ref{eq:excitonmodel1} used in the main text, we first note that, by wannierizing the LUMO band Bloch functions as 
\begin{align}
\xi_{c_1 \ve{R'}}(\ve r) =\frac{1}{\sqrt{N_q}}\sum_{\ve q}^{\mathrm{BZ}}e^{-\mathrm{i}\ve q \cdot \ve R'} \xi_{c_1\ve q}(\ve r),
\end{align}
we can also write the Dyson orbital as a linear combination of localized functions, which span the same space as the Bloch functions
\begin{align}
  \label{eq:wannier_dyson3}
  &D_{v_1 \ve{R}=0}^{m}(\ve r) =
  \sum_{\ve{R'}}^{\mathrm{crystal}} c_{v_1\ve{R'}}^{m} \xi_{c_1\ve{R'}}(\ve r),
\end{align}
where the coefficients $c_{v_1\ve{R'}}^{m}$ are connected to the BSE eigenvector via the inverse Fourier transform (multiplied by an additional $1/\sqrt{N_{q}}$ factor)
\begin{align}
  \label{eq:BSE_evec_FT}
  &c_{v_1\ve{R'}}^{m} =
  \frac{1}{N_{q}}\sum_{\ve{q}}^{\mathrm{BZ}} X_{v_1 c_1 \ve{q}}^{m} e^{\mathrm{i} \ve{q}\cdot\ve{R'}}.
\end{align}
Next, we approximate the generic coefficients $c_{v_1\ve{R'}}^{m}$ in Eq.~\Ref{eq:wannier_dyson3} by the product of a real-valued envelope function $\alpha(\ve{R})$, controlling the extent of the Dyson orbital, and a site-dependent complex phase factor $e^{\mathrm{i} \beta(\ve{R})}$.
Finally, we approximate the wannierized LUMO functions $\xi_{c_1\ve{R'}}$ in Eq.~\Ref{eq:wannier_dyson3} by the LUMO of an isolated, single molecule calculation $\xi_{L}$, placed at the lattice sites $\ve{R'}$ and arrive at the model (Eq.~\ref{eq:excitonmodel1} in the main text)
\begin{align}
  \label{eq:excitonmodel1_si}
  &D_{v_1 \ve{R}=0}^{m}(\ve r) =
  \sum_{\ve{R'}}^{\mathrm{crystal}} \alpha(\ve{R'}) e^{\mathrm{i} \beta(\ve{R'})} \xi_{L}(\ve r - \ve{R'}).
\end{align}

Via the connection Eq.~\ref{eq:BSE_evec_FT} (see also Fig.~\ref{sifig5}b,f and Fig.~\ref{sifig6}b,f) between the BSE eigenvector and the model coefficients, it becomes clear that we not only can use our model wave function to simulate photoemission from an excited state, but can also employ it to gain understanding of the structure of excited states (e.g. by fitting the model to experimental data and using Eq.~\ref{eq:BSE_evec_FT} to construct the exciton wave function).
Conversely, we can use the exciton wave function from an \emph{ab initio} calculation to motivate a model choice ($\alpha$ and $\beta$), which can then be used for systems or excitations, where the \emph{ab initio} data may not be available.
For the present case of a 6T monolayer, using a Gaussian envelope function $\alpha$ proved successful.
For $\beta$, we propose two choices: 
(i) Setting $e^{\mathrm{i} \beta(\ve{R})}=e^{\mathrm{i} \ve{q} \ve{R}}$, the phase changes periodically across the surface, leading to a constant phase difference between neighboring molecules. This resembles exactly the situation found in $S_1$, where the phase difference between neighboring molecules was found to be $\sim  195^{\circ}$, see Fig.~\ref{fig3}c.
(ii) Allowing $\beta(\ve{R})$ to only be $0$ or $\pi$, we get $e^{\mathrm{i} \beta(\ve{R})}=\pm 1$. At each molecular site, the value of $\beta$ may be chosen individually, such that two neighbors may either have equal phases, or a phase difference of $\pi$. This structure is found in $S_3$, see Fig.~\ref{sifig2}b, and to some approximation also works for modeling $S_1$ (since $195^{\circ}$ is not too far from a phase change of $\pi$). Based on the $\mathrm{\pi}$-orbital structure of the LUMO wave function, this model choice can be understood as a combination of antibonding (no phase change) and bonding (phase change of $\pi$) overlaps, as can be seen in Figs.~\ref{fig3}B and \ref{sifig2}A.

As shown in Eq.~\ref{eq:expotperiodic_6T1}, the photoemission intensity for $S_1$ and $S_3$ of the 6T monolayer system is simply the absolute square of a Fourier-transformed Wannier-Dyson orbital. Using our model function Eq.~\ref{eq:excitonmodel1_si} for the Wannier-Dyson orbital, we thus find
\begin{equation}
\label{eq:excitonmodel2_si}
I_\mathrm{model}(\ve{k}) \propto \left| \sum_{\ve{R}}^{\mathrm{crystal}} \alpha(\ve{R}) e^{\mathrm{i} \beta(\ve{R})} e^{-\mathrm{i} \ve{k} \cdot\ve{R}} \right|^2 \cdot \left| \mathcal{F} \left[ \xi_L \right] (\ve{k}) \right|^2.
\end{equation} 
This equation corresponds to Eq.~\ref{eq:excitonmodel2} in the main text. 
As is shown in Sec.~S3.3 below, this simple form of the photoemission intensity as a product of two terms allows for direct insights into the origin of fine details found in momentum maps.

To summarize a lengthy derivation, we have thus shown that the model used to analyze our experimental data can be derived from the accurate equations of the exPOT formalism. 

For comparability, we conclude this section by bringing the \emph{ab initio} photoemission expression Eq.~\ref{eq:expotperiodic_6T1} to a form similar to the model expression Eq.~\ref{eq:excitonmodel2_si}. 
First, we insert Eq.~\ref{eq:wannier_dyson2} into Eq.~\ref{eq:expotperiodic_6T1} to get
\begin{align}
  I_{m}(\ve k) \propto &
  \left|\mathcal F \left[\sum_{\ve{q'}}^{\mathrm{BZ}} X_{v_1 c_1 \ve{q'}}^{m} \xi_{c_1\ve{q'}}\right](\ve{q} +\ve G)\right|^2, \nonumber \\
  & \mathrm{with}~\ve q + \ve G = \ve k.
\end{align}
Then, we use the linearity of the Fourier transform and momentum conservation (see text below Eq.~\ref{eq:ft_of_bloch}) to arrive at the final expression (Eq.~\ref{eq:exPOTperiodicsimple} in the main text)
\begin{align}
  \label{eq:expotperiodic_6T2}
  I_{m}(\ve k) \propto &
  \left|X_{v_1 c_1 \ve{q}}^{m}\right|^2  \cdot
  \left|\mathcal F \left[\xi_{c_1\ve{q}}\right](\ve{q} +\ve G)\right|^2, \nonumber \\
  & \mathrm{with}~\ve q + \ve G = \ve k.
\end{align}

\subsection{Comparison of photoemission from \emph{ab initio} calculations and the exciton model}\label{sec:comp_abinitio_model}

\begin{figure}[h]
\centering
\includegraphics[width=1.0\textwidth]{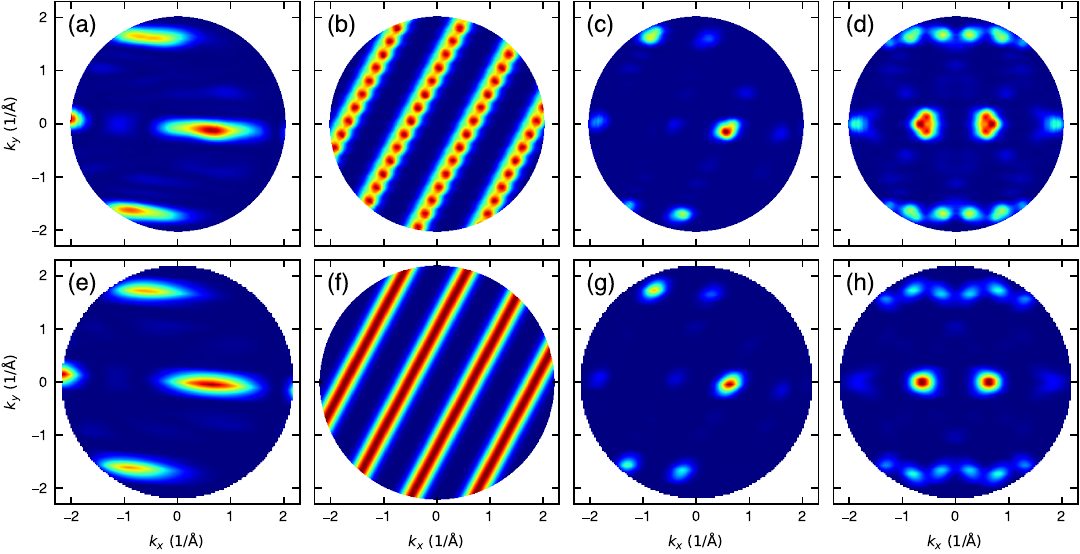}
\caption{
(a)--(d) Decomposition of the photoemission intensity of exciton $S_1$ from the \emph{ab initio} exPOT simulation into individual contributions.  (a) Absolute square of the Fourier-transformed LUMO band wave functions $\left|\mathcal F \left[\xi_{c_1\ve{q}}\right](\ve{q} +\ve G)\right|^2$. (b) Absolute square of the BSE eigenvector $\left|X_{v_1 c_1 \ve{q}}^{m}\right|^2$ from  Eq.~\ref{eq:expotperiodic_6T2}. Panel (d) shows the symmetrized photoemission intensity shown in panel (c).
(e)--(h) Analogous to (a)--(d) but for the exciton model following Eq.~\ref{eq:excitonmodel2_si}. (e) Absolute square of a Fourier-transformed LUMO of a gas-phase molecule $\left| \mathcal{F} \left[ \xi_L \right] (\ve{k}) \right|^2$. (f) Factor $| \sum_{\ve{R}}^{\mathrm{crystal}} \alpha(\ve{R}) e^{\mathrm{i} \beta(\ve{R})} e^{-\mathrm{i} \ve{k} \cdot\ve{R}} |^2$ arising from the envelope function and model phase factor.  Panel (g) and (h) show the photoemission intensity $I_\mathrm{model}(\ve{k})$ (product of panels (e) and (f)), before and after symmetrization, respectively.}\label{sifig5}
\end{figure}

The similar form of the model photoemission expression Eq.~\ref{eq:excitonmodel2_si} and the \emph{ab initio} expression Eq.~\ref{eq:expotperiodic_6T2} invites for a direct comparison of the two approaches.
To start, Fig.~\ref{sifig5} shows a comparison of a photoemission map of $S_1$ from the monolayer \emph{ab initio} calculation (panels (a)--(d)), with the corresponding photoemission simulation based on the model depicted in panels (e)--(h).

For the \emph{ab initio} simulation, based on Eq.~\ref{eq:expotperiodic_6T2}, we can simply calculate the photoemission intensity $I_{m}(\ve k)$, panel (c), as the product of the LUMO band photoemission map $\left|\mathcal F \left[\xi_{c_1\ve{q}}\right](\ve{q} +\ve G)\right|^2$, panel (a), and the BSE-eigenvector $\left|X_{v_1 c_1 \ve{q}}^{m}\right|^2$, panel (b).
We note that the LUMO band map trivially shows the tilt (Fig.~\ref{sifig1}b) and the small misalignment of the long molecular axis with respect to the $a_1$ axis (Fig.~\ref{sifig1}a) of the monolayer geometry.
Also note that the BSE eigenvector exhibits a stripe-like structure, with the angle of the stripes being determined by the unit-cell slip, see Fig.~\ref{sifig1}.
Panel (d) shows the symmetrized (for 6T molecules on the Cu(110)-$(2 \times 1)$O substrate) photoemission intensity, exhibiting the now well-known double-peak along the $k_y=0$ line.

In comparison, the second row of Fig.~\ref{sifig5} shows the photoemission intensity calculated for the model wave function Eq.~\ref{eq:excitonmodel1_si}, that is, based on Eq.~\ref{eq:excitonmodel2_si}.
For the LUMO wave function $\xi_L$ ($\left| \mathcal{F} \left[ \xi_L \right] (\ve{k}) \right|^2$ in panel (e), we use the LUMO from a DFT calculation of an isolated 6T molecule and rotate it in accordance with the monolayer geometry. 
Panel (f) shows the first factor on the right hand side of Eq.~\ref{eq:excitonmodel2_si}, $| \sum_{\ve{R}}^{\mathrm{crystal}} \alpha(\ve{R}) e^{\mathrm{i} \beta(\ve{R})} e^{-\mathrm{i} \ve{k} \cdot\ve{R}} |^2$, and is thus determined by the envelope function and phase factor used in the model. The molecular sites $\ve R$ are taken from the monolayer geometry and we use a Gaussian envelope function $\alpha$ with a $\mathrm{FWHM}_x$ of $13$~{\AA}, in reasonable agreement with the estimate of $16$~{\AA} given above, and phase factors $e^{\mathrm{i} \beta(\ve{R})}$ corresponding to a $195^{\circ}$ phase change between neighboring molecules (in $\ve{\mathrm{a_2}}$ direction).
The resulting factor then shows the same stripe-like nature as the BSE eigenvector, making the above-shown link, Eq.~\ref{eq:BSE_evec_FT}, between the two quantities obvious. 
Finally, panels (g) and (h) again show the unsymmetrized and symmetrized photoemission intensities $I_\mathrm{model}(\ve{k})$, respectively. Again, comparison to the \emph{ab initio} simulations in panels (c) and (d) shows great resemblance, demonstrating the strength of the model, allowing us to understand most of the details of $S_1$ in terms of just the used $\mathrm{FWHM}_x$ and phase change.

\begin{figure}[h]
\centering
\includegraphics[width=1.0\textwidth]{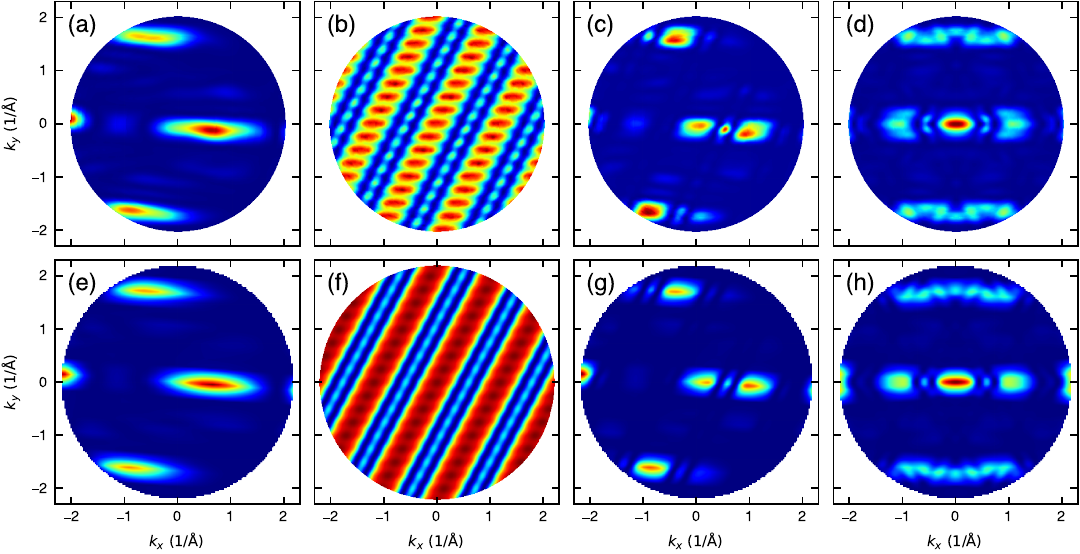}
\caption{Analogous to Fig.~\ref{sifig5} but for exciton $S_3$, the first row shows \emph{ab initio} photoemission as calculated with Eq.~\ref{eq:expotperiodic_6T2}, while the second row shows photoemission using the model Eq.~\ref{eq:excitonmodel1_si} and is thus based on Eq.~\ref{eq:excitonmodel2_si}.}\label{sifig6}
\end{figure}

Similarly, Fig.~\ref{sifig6} shows photoemission from $S_3$, with the first row again showing the \emph{ab initio} exPOT result and the second row showing photoemission based on the model.
For the model, we use a Gaussian envelope function $\alpha$ with a $\mathrm{FWHM}_x$ of $18$~{\AA}, which is again in good agreement with the $19.9$~{\AA} estimate given before, and assign phase factors $e^{\mathrm{i} \beta(\ve{R})} = \pm 1$ based on Fig.~\ref{sifig2}b.
Comparing panels (b) and (f), we can see that the model reproduces both the major and minor stripes of the BSE eigenvector.
This leads to very good agreement of the symmetrized photoemission intensities shown in D and H, where also the minor-lobe structure agrees very well.
Thus, the excited state photoemission can be reproduced to high accuracy by the model Eq.~\ref{eq:excitonmodel1_si}, and can therefore be understood via a combination of bonding and antibonding LUMO wave functions weighted by a Gaussian envelope function.

%

\end{document}